\documentclass[preprint]{egpubl}
\usepackage{WICED2022}
 
\WsSubmission      

\usepackage[T1]{fontenc}
\usepackage{dfadobe}  

\usepackage{cite}  
\BibtexOrBiblatex
\electronicVersion
\PrintedOrElectronic
\usepackage{graphicx}

\usepackage{egweblnk} 

\usepackage{listings}
\usepackage{tikz}
\usepackage{tikz-qtree-compat}
\usepackage{subfigure}
\usepackage{caption} 
\usepackage{multirow}

\title[The Prose Storyboard Language Version 2.0, Revised and Illustrated Edition]{The Prose Storyboard Language:\\  A Tool for Annotating and Directing Movies\\ (Version 2.0, Revised and Illustrated Edition)}
\author[Ronfard et al.]
{\parbox{\textwidth}{\centering Rémi Ronfard$^{1}$\orcid{0000-0003-4830-5690}
        and Vineet Gandhi$^{2}$\orcid{0000-0001-8861-7731} 
        and Laurent Boiron$^{3}$
        and Vaishnavi Ameya Murukutla$^{1}$\orcid{0000-0001-9455-3391} 
        }
        \\
{\parbox{\textwidth}{\centering $^1$Univ. Grenoble Alpes, Inria, CNRS, Grenoble INP, LJK, France\\
         $^2$International Institute of Information Technology, Hyderabad, India\\
         $^3$Weta Digital, New Zeland
       }
}
}

\begin{document}
\maketitle

\begin{abstract}
The prose storyboard language is a formal language for describing movies shot by shot, where 
each shot is described with a unique sentence.  The language uses a simple syntax and limited vocabulary 
borrowed from working practices in traditional movie-making and is intended to be readable both by machines
and humans. The language has been designed over the last ten years to serve as a high-level user interface 
for intelligent cinematography and editing systems. In this new paper, we present the latest evolution of
the language,  and the results of an extensive annotation exercise showing the benefits of the language
in the task of annotating the sophisticated cinematography and film editing of classic movies. 


\end{abstract}

\begin{figure*}
\centering
        \includegraphics[width = 0.32\linewidth]{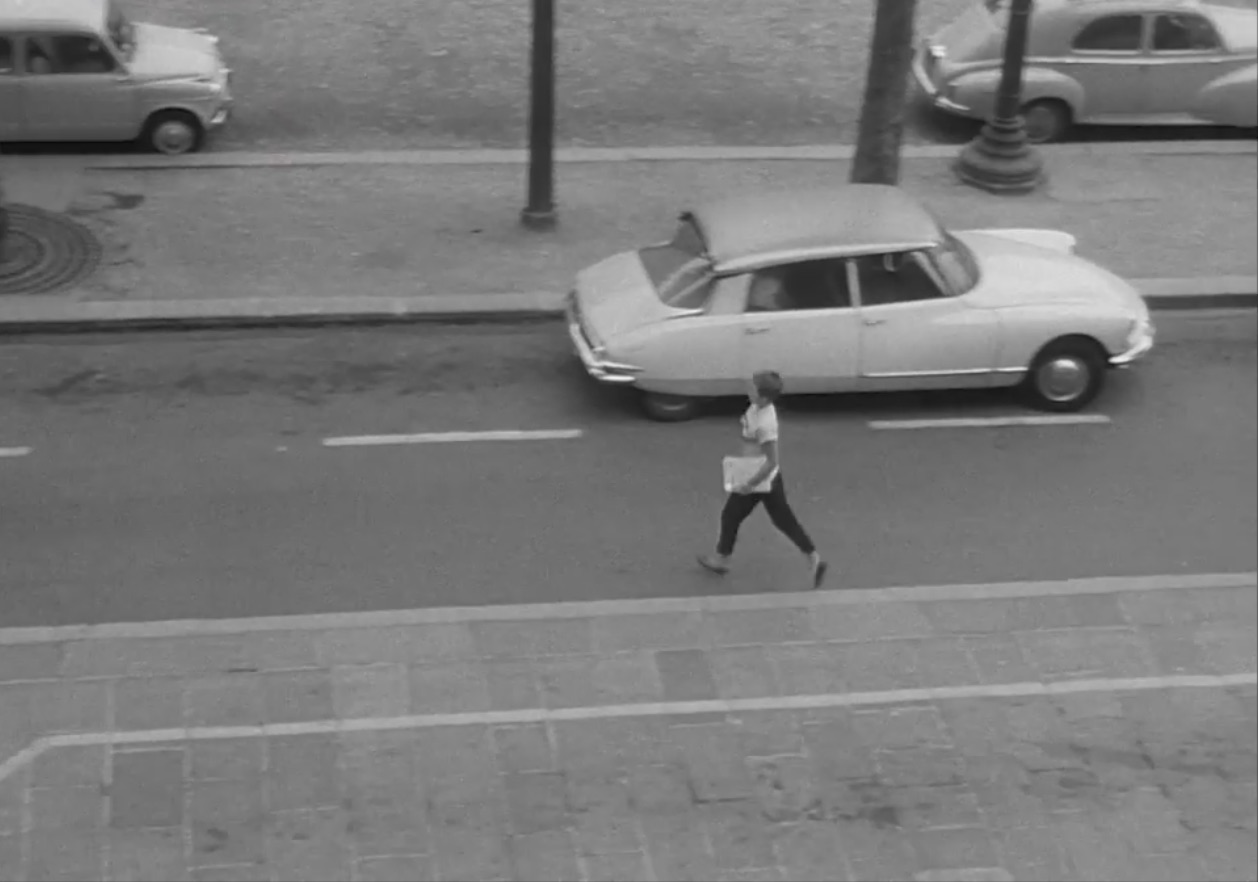}
        \includegraphics[width = 0.32\linewidth]{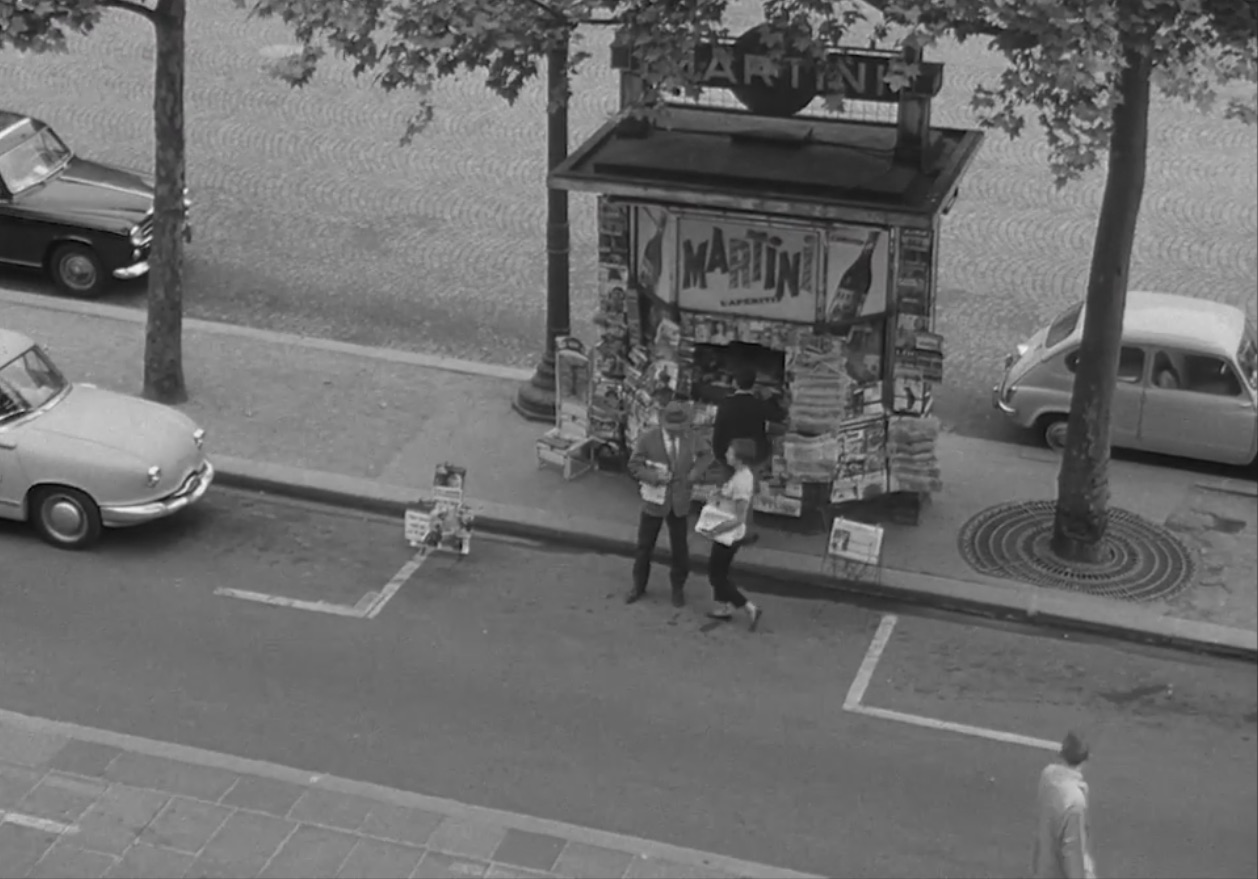}
        \includegraphics[width = 0.32\linewidth]{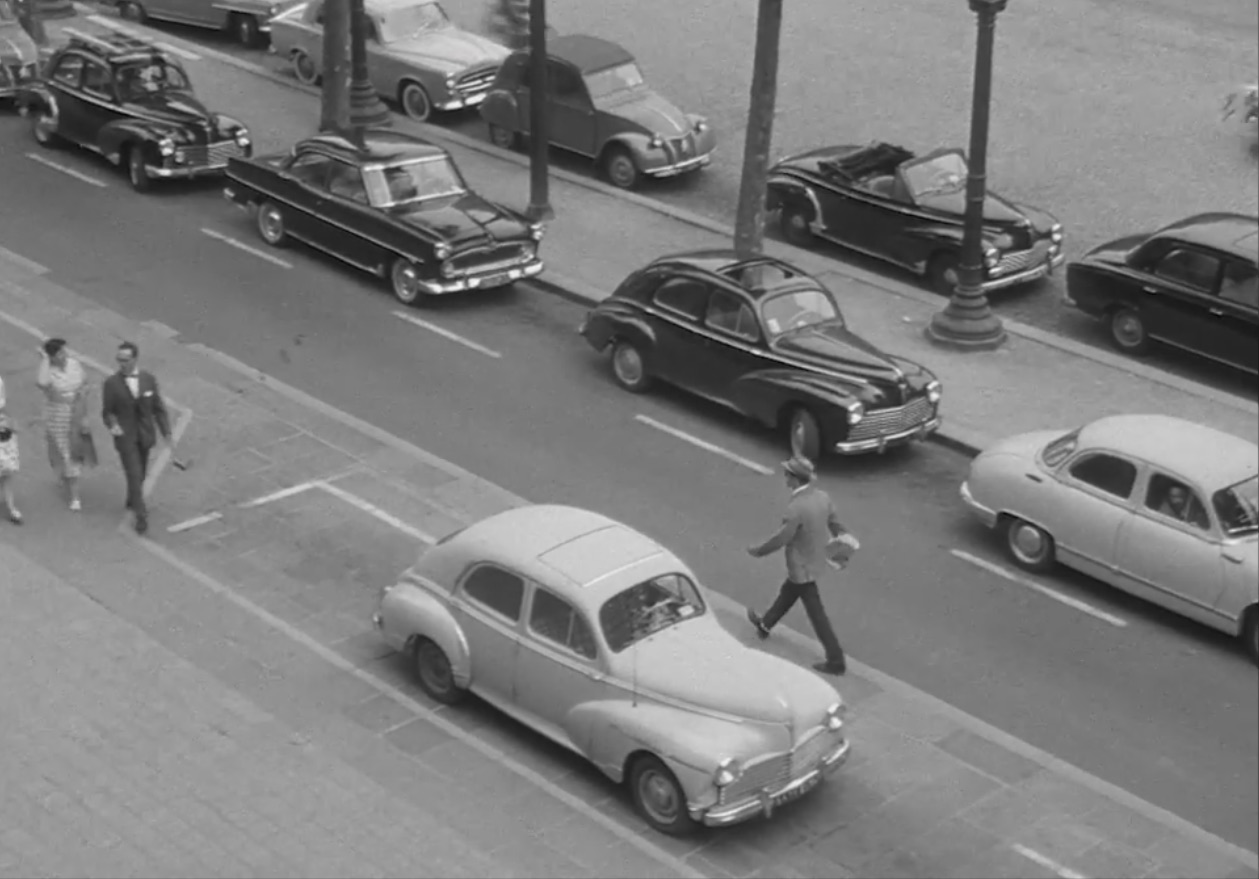}
\begin{tikzpicture}[grow'=up,scale=0.6]
\tikzset{every tree node/.style={align=center,anchor=base}}
\tikzset{level 1+/.style={level distance=2\baselineskip}}
 \tikzset{frontier/.style={distance from root=8.5\baselineskip}}
\Tree [.Shot [\qroof{high angle LS Pat 34backleft}.Composition ] [\qroof{while Pat moves to screen left}.Event ]  [. Recomposition [\qroof{then pan left to}.CameraTo ] [\qroof{high angle LS Mi Pat left }.Composition ]] [. Recomposition [\qroof{then as Mi moves to screen left}.Cue ]  [\qroof{pan left to}.CameraTo ] [\qroof{high angle LS Mi 34backleft }.Composition ]]]
\end{tikzpicture}
\caption{Complex shot in \emph{Breathless}}
\label{fig:Breathless}
\end{figure*}

\section{Introduction}
In movie production, directors often use a semi-formal idiom of natural language to convey the shots they 
want to their cinematographer.  Similarly, film scholars use a semi-formal idiom of natural language to describe the visual composition of shots in produced movies to their readers. In order to build intelligent and expressive virtual cinematography and editing systems, we believe the same kind of high-level descriptions need to be agreed upon.  In this paper, we propose a formal language that can serve that role.  Our primary goal in proposing this language is to build software cinematography agents that can take such formal descriptions as input, and produce fully realized shots as an output. A secondary goal is to perform in-depth studies of film style by automatically analyzing real movies and their scripts in terms of the proposed language.

The prose storyboard language is a formal language for describing shots visually. We leave the description of the soundtrack for future work. The prose storyboard language separately describes the spatial structure of individual movie frames (compositions) and their temporal structure (shots). In film analysis, there is frequent confusion between shots and compositions. A {\em medium shot} describes a composition, not a shot. If the actor moves towards the camera in the same shot, the composition will change to a {\em close shot} and so on. Therefore, a general language for describing shots cannot be limited to describing compositions such as {\em medium shot} or {\em close shot} but should also describe screen events which change the composition during the shot.  In the prose storyboard language, each shot is a complete sentence with at least one composition and an arbitrary number of screen events. This offers unprecedented expressive power for describing and directing movies.


Our language can be used indifferently to describe shots in pre-production (when the movie only exists in the screen-writer and director's minds), during production (when the camera records a continuous "shot" between the times when the director calls "camera" and "cut"), in post-production (when shots are cut and assembled by the film editor) or to describe existing movies.  The description of an entire movie is an ordered list of sentences, one per shot.  Exceptionally, a movie with a single shot, such as Rope by Alfred Hitchcock, can be described with a single, long sentence.

In this paper, we assume that all shot descriptions are manually created. We leave for future work the important issue of automatically generating prose storyboards from existing movies, where a number of existing techniques can be used \cite{Brand97,DMR05,GoldmanCSS06,GR13}.  We also leave for future work the difficult problems of automatically generating movies from their prose storyboards, where existing techniques in virtual camera control can be used \cite{He96,Christianson96,Jhala06,CO06,Galvane2013,Galvane2014,Galvane2015}.

\section{Prior art}
Our language is loosely based on existing practices in movie-making \cite{Thompson09a,Thompson09b} and previous research in the history of film style \cite{Bordwell98,Salt09}. Our language is also related to the common practice of graphic storyboards. In a graphic storyboard, each composition is illustrated with a single drawing. The blocking of the camera and actors can be depicted with a conventional system of arrows within each frame, or with a separate set of floor plan views, or with titles between frames.

In our case, the transitions between compositions use a small vocabulary of screen events including camera actions  (pan, dolly, crane, hold, continue)  and actor actions (speak, react, move, cross). Although the action vocabulary could easily be extended, we voluntarily keep it small because our focus in this paper is restricted to the blocking of actors and cameras, not the high-level semantics of the narrative.

We borrow the term prose storyboard from Proferes \cite{Proferes08} who used it as a technique for decomposing a films script into a sequence of shots, expressed in natural language. Other authors use the French term "decoupage" to describe this important step in film production. The name catches the intuition that the language should directly translate to images. In contrast to Proferes, our prose storyboard language is a formal language, with a well-defined syntax and semantics, suitable for future work in intelligent cinematography and editing.  

Our proposal is complementary to the Movie Script Markup Language (MSML) \cite{RijsselbergenKVMW09}, which encodes the structure of a movie script.  
In MSML, a script is decomposed into dialogue and action blocks but does not describe how each block is translated into shots. Our prose storyboard language can be used to describe the blocking of the shots in a movie in relation to an MSML-encoded movie script. For this purpose, MSML makes provision for a Manufacturing model and Animation model, which are only loosely described. The prose storyboard language can be seen as an alternative representation for both the Manufacturing and Animation models in MSML. 

Our proposal is also related to the Declarative Camera Control Language (DCCL) which describes film idioms, not in terms of cameras in world coordinates but in terms of shots in screen coordinates \cite{Christianson96}. The DCCL is compiled into a film tree, which contains all the possible editings of the input actions, where actions are represented as subject-verb-object triples. Our prose storyboard language can be used in coordination with such constructs to guide a more extensive set of shot categories, including complex and developing shots. 

Our approach is also related to the work of Jhala and Young who used the movie Rope by Alfred Hitchcock to demonstrate how the storyline and the director's goal should be represented to an automatic editing system \cite{Jhala06}. They used Crossbow, a partial order causal link planner, to solve for the best editing, according to a variety of strategies, including maintaining tempo and depicting emotion. They demonstrated the capability of their solver to present the same sequence in different editing styles. But their approach does not attempt to describe the set of possible shots. Our prose storyboard language attempts to fill that gap.  

Other previous work in virtual cinematography \cite{Shen03, Jhala05, Friedman06, ONeillRN09, MarkowitzKSB11,Galvane2014,Galvane2015,Leake2017} has been limited to simple shots with either a static camera or a single uniform camera movement. Our prose storyboard language is complementary to such previous work and can be used to define higher-level cinematic strategies, including arbitrarily complex combinations of camera and actor movements, for most existing virtual cinematography systems.

Text-to-movie (or text-to-scene) authoring is a general class of methods that have been proposed for automatically generating 3D graphics and animation from natural language text. Good results have been obtained in limited domains, such as generating 3D scenes from natural language accident reports~\cite{Akerberg2003} or generating cartoon animation from scripted dialogue scenes~\cite{Seversky2006}.  Commercially available text-to-movie systems include  NawmalMAKE  
and Plotagon Studio 
Such systems use marked-up dialogues as input and generate simple shots with minimal camera movements and editing. Other related work along the same lines includes Ye and Baldwin \cite{Ye2008} who describe a method for generating storyboards from movie scripts; Marti et al. \cite{Marti2018} who describe methods for generating previz animation, also from movie scripts. 
 
Closer to our approach, Director Notation \cite{Yannopoulos2013} is a symbolic language intended to express the content of film (motion pictures), much as musical notation provides a language for the writing of music. But DN is a graphical notation whereas PSL is a pseudo-natural language, and DN describes the movie production process, whereas PSL describes the movie itself, as in a storyboard. TIMISTO \cite{Vogt2013} and SLAP \cite{Braga2016} are  pattern languages for creating animation from storyboards.


A previous version of the prose storyboard language was presented at the international workshop on intelligent cinematography and editing (WICED) in 2012. Since then, several variations of PSL have been used for generating cinematic replays in serious games \cite{Galvane2014}, for generating synthetic complex shots from live video material \cite{GR14,GR15}, for staging complex scenes in 3D animation \cite{Louarn18}, for directing cinematographic drones \cite{Galvane2018} and for learning film editing patterns from examples \cite{Wu2018}. In this revised edition, we provide a definitive version of the language, illustrated with a large number of examples and a reference implementation, in the hope that it will stimulate future work in automatic annotation and generation of movies.

\begin{figure}
\centering
\includegraphics[width=0.49\columnwidth]{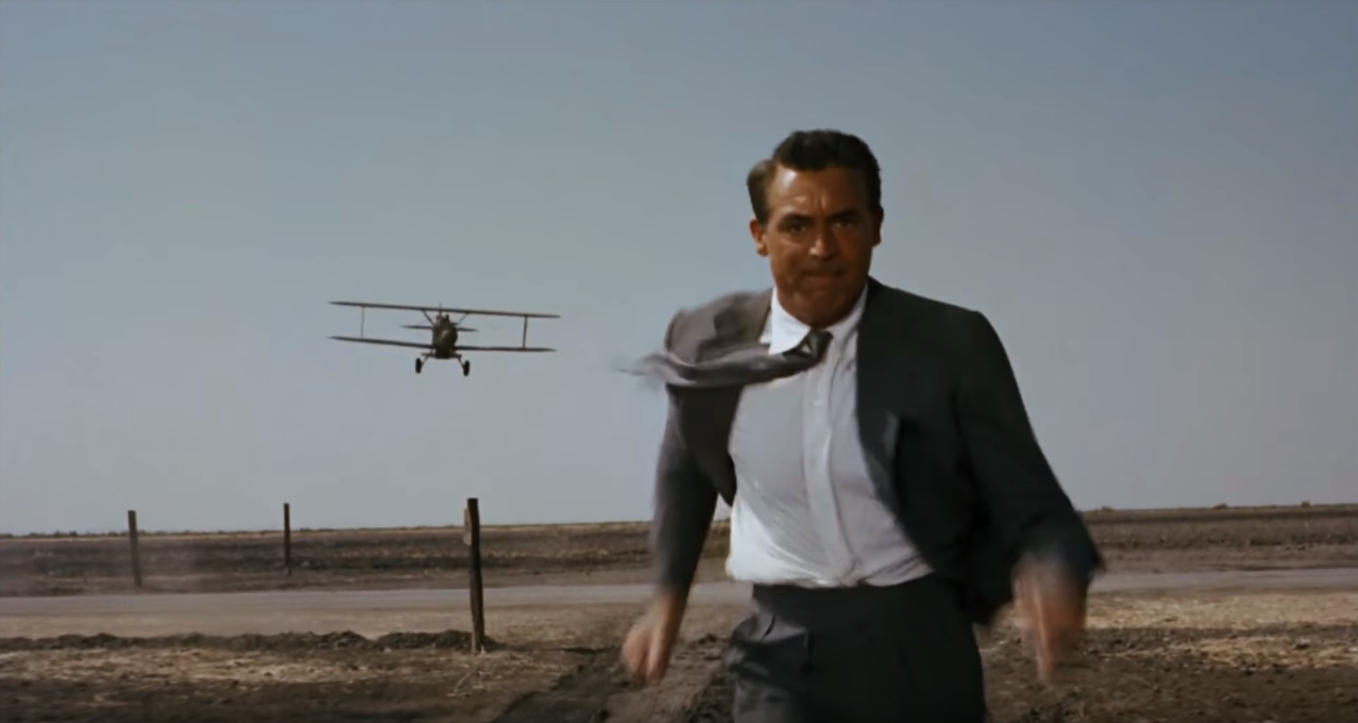}
\includegraphics[width=0.49\columnwidth]{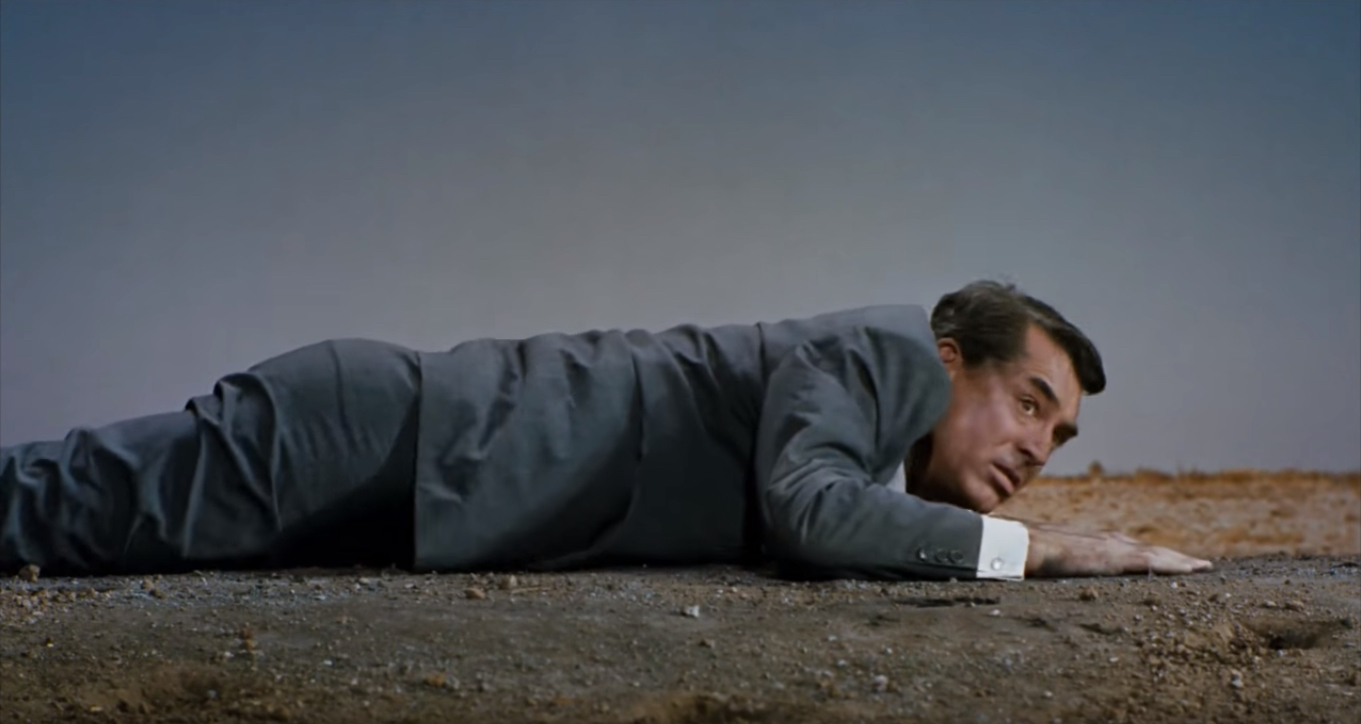}
\begin{tikzpicture}[grow'=up,scale=0.95]
\tikzset{every tree node/.style={align=center,anchor=base}}
\tikzset{level 1+/.style={level distance=2\baselineskip}}
 \tikzset{frontier/.style={distance from root=5.5\baselineskip}}
  \Tree [.Scene [.Shot [\qroof{low angle FS Plane MS Th front}.Composition ] ][.Shot [\qroof{cut to}.Transition ][\qroof{MLS Th 34right}.Composition ] ] ]
\end{tikzpicture}
\caption{Prose storyboard language description of two iconic shots in 
Alfred Hitchcock's \emph{North By Northwest}}
\label{fig:CD_shots}
\end{figure}

\section{Requirements}
The prose storyboard language is designed to be expressive, i.e. it should describe arbitrarily complex shots, while at the same time being compact and intuitive.  Our approach has been to keep the description of simple shots as simple as possible, and allowing for more complex descriptions when needed. Thus, for example, we describe actors in composition from left to right, which is an economical and intuitive way of specifying relative actor positions in most cases. As a result, our prose storyboard language is very close to natural language (see Fig.\ref{fig:rope-frames} for examples). 

It should be easy to parse the language into a non-ambiguous semantic representation that can be matched to video content, either for the purpose of describing existing content or for generating novel content that matches the description. It should therefore be possible (at least in theory) to translate any sentence in the language into a sketch storyboard, then to a fully animated sequence. 

It should also be theoretically possible to translate existing video content into a prose storyboard. This puts another requirement on the language, that it should be possible to describe existing shots just by watching them.  There should be no need for contextual information, except for place and character names. As a result, the prose storyboard language can also be used as a tool for annotating complete movies and for logging shots before post-production. Since the annotator has no access to the details of the shooting plan, even during post-production \cite{Murch86,Ondaatje04}, we must therefore make it possible to describe the shots in screen coordinates, without any reference to world coordinates. 


\section{Syntax and semantics}
The prose storyboard language is a context-free language, whose terminals include generic and specific terms. Generic terminals are used to describe the main categories of screen events including camera actions (pan, dolly, cut, dissolve, etc.) and actor actions (enter, exit, cross, move, speak, react, etc.). Specific terminals are the names of characters, places and objects that compose the image and play a part in the story. Non-terminals of the language include important categories of shots (simple, complex, developing), image compositions and developments. The complete grammar for the language is illustrated with the AND/OR graph in Fig. \ref{fig:psl_andortree} and described in the PEG notation in Fig.\ref{grammar}. A reference implementation using the Parsimonious toolkit  for parsing PSL sentences using the Python language can be found at \url{https://gitlab.inria.fr/vmurukut/psl}.


The semantics of the prose storyboard language is best described in terms of a Timed Petri Net (TPN) where durative events such as compositions and camera actions are represented as  {\em places} ; and instantaneous events (such as cuts and the start and end of other events) are represented as {\em transitions}. We use timed Petri nets, rather than finite state machines, as a semantic representation of shots in a prose storyboard, in order to adequately represent developing shots with an elaborate choreography of actor and camera movements taking place simultaneously, such as the opening shot in Orson Welles' "Touch of evil". 

TPNs  have been proposed for representing the temporal structure of movie scripts \cite{RijsselbergenKVMW09}, character animation \cite{Magalhaes98, Blackwell01}, game authoring \cite{BalasBAG08}, turn-taking in conversation \cite{Chao12} and synchronisation and storage models for multimedia systems \cite{Little1990SynchronizationAS}. Our model differs from previous work by representing all durative events with places and using transitions to synchronize them.  For lack of space, we differ the description of the Petri Net interpretation of the prose storyboard language to future work. 


\begin{figure}
    \centering
    \subfigure[Shot sizes]
    {
      \includegraphics[width=0.8\columnwidth]{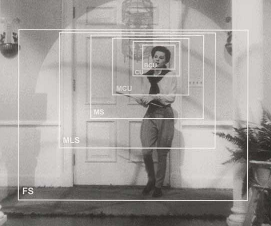}
      \label{fig:shottype}
    }
    \subfigure[Profile angle]
    {
    \includegraphics[width=0.8\columnwidth]{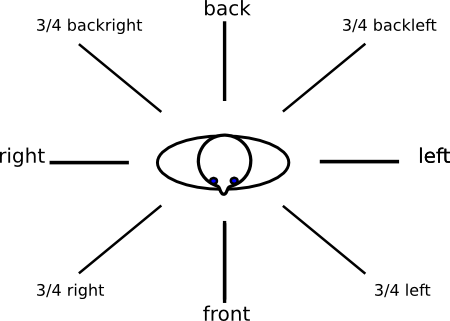}
    \label{fig:profileangle}
    }
    \caption{\subref{fig:shottype} shows shot sizes in the prose storyboard (reproduced from \protect\cite{Salt06}). \subref{fig:profileangle} shows the profile angle of an actor defines his orientation relative to the camera. For example, an actor with a  \emph{left} profile angle is oriented with his left side facing  the camera. }
\end{figure}

\begin{figure}
    \centering
    \subfigure[Basis for shot size]
    {
    \includegraphics[width=0.8\columnwidth]{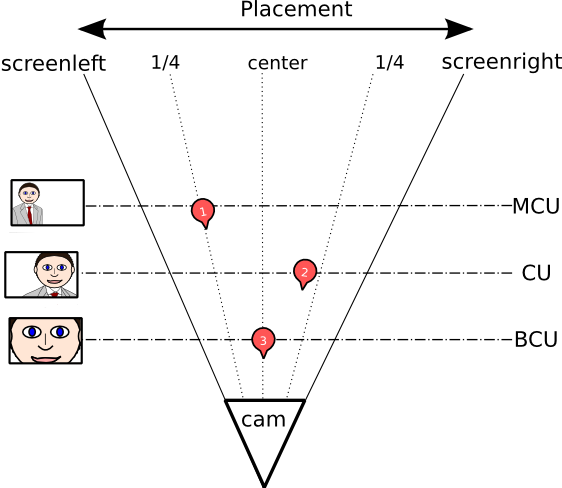}
    \label{fig:compocoord}
    }
    \subfigure[Screen coordinates]
    {
    \includegraphics[width=0.8\columnwidth]{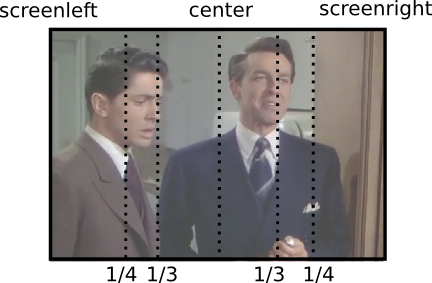}
    \label{fig:placement}
    }
    \caption{Shot size is a function of the distance between  the camera and actors,as well as  the camera focal length,as seen in \subref{fig:compocoord}. \subref{fig:placement} shows the horizontal placement of actors in a composition is expressed in screen coordinates.} 
\end{figure}

\begin{figure}
    \centering
    \subfigure [low angle ECU Girl 34left]
    {
        \includegraphics[width=0.8\columnwidth]{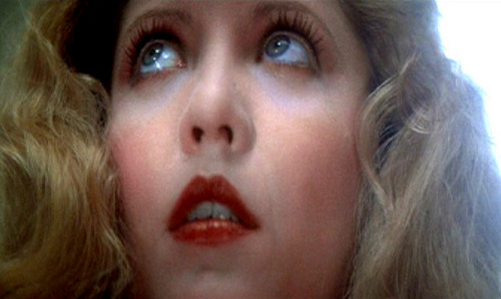}
        \label{fig:sub1}
    }
    \caption{Single actor composition in \emph{Prose Storyboard Language}. This frame from Brian De Palma's \emph{Dressed To Kill}(1980) shows an extreme close up shot from a low angle.}

    \subfigure[FS Cyd 34right Fred front]
    {
        \includegraphics[width=0.8\columnwidth]{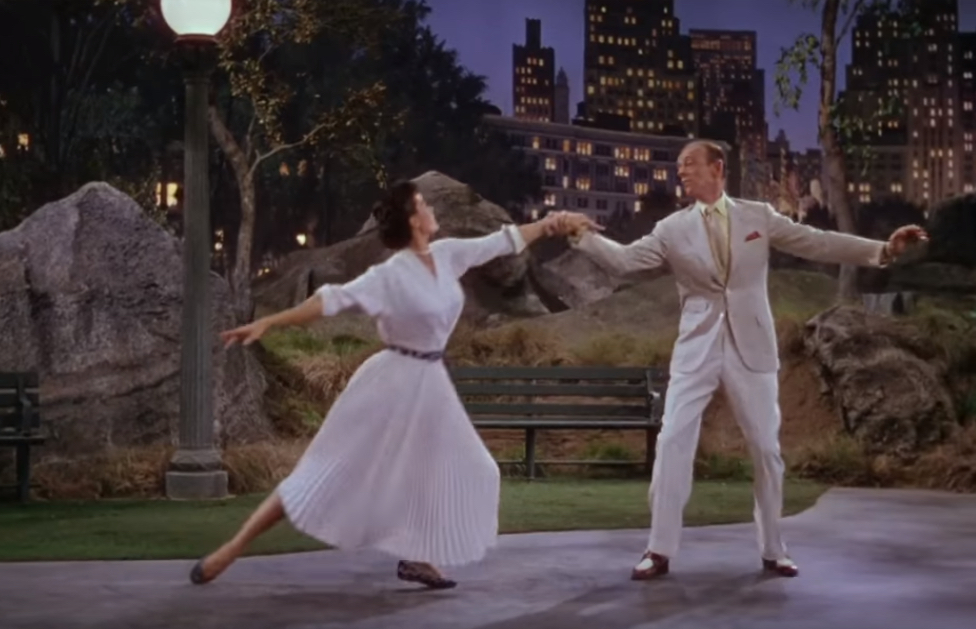}
        \label{fig:sub2}
    }
    \subfigure[MS Girl 34backright Ferdinand front]
    {
        \includegraphics[width=0.8\columnwidth]{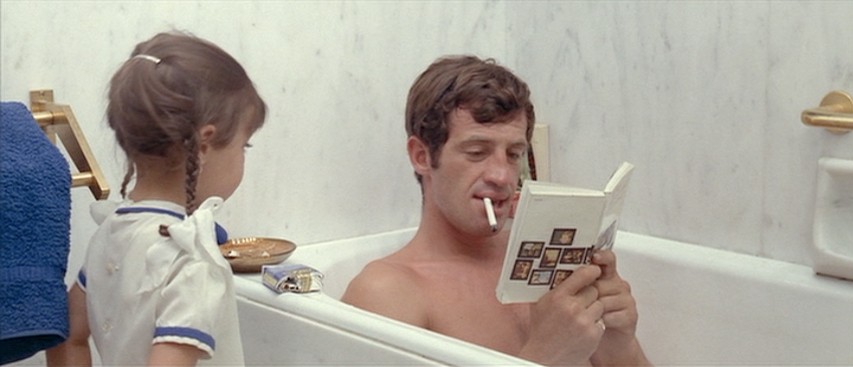}
        \label{fig:sub3}
    }
    \caption{Two actor composition in \emph{Prose Storyboard Language}. Compositions from Vincente Minnelli's 1953 musical, \emph{The Band Wagon} ~\subref{fig:sub2} and Jean-Luc Godard's 1965 French New Wave film, \emph{Pierrot Le Fou} ~\subref{fig:sub3} feature two actors in a frame at different sizes.}
\end{figure}

\section{Image Composition}

Image composition is the way to organise visual elements in the motion picture frame to deliver a specific message to the audience. In our work, we propose a formal way to describe image
composition in terms of the actors and objects present on the screen and the spatial and temporal relations between them.

Following Thomson and Bowen \cite{Thompson09a}, we define a composition as the relative position and orientation  of visual elements called \emph{Subject}, in screen coordinates. In the simple case of {\em flat staging}, all subjects are more or less in the same plane with the same size, but in the case of {\em deep staging}, different subjects are seen at different  sizes, in different planes. For the sake of generality, we therefore choose to indicate the size of each subject separately. As a result, each subject is defined by its size, profile angle, and screen position. See Fig. \ref{fig:profileangle} and \ref{fig:profileangle} for a visual explanation.

As a convention, we assume that the subjects are described from left to right. This means that the left-to-right ordering 
of actors and objects is part of the composition. Indeed, because the left-to-right ordering of subjects is so important in 
cinematography and film editing, we introduce a special keyword {\em cross}  for the screen event of one subject crossing over or under another subject. Shot sizes are used to describe the relative sizes of subjects independently of the camera lens as illustrated in Fig.\ref{fig:compocoord}.


The $Screen$ term describe the subject position in screen coordinate. It allows to slightly modify the generic framing by shifting the subject position to the left or right corner as shown in Fig.\ref{fig:placement}. Default  $Screen$ values are used to describe symetric compositions where subjects are evenly distributed from left to right. Non-default  $Screen$  values are used to describe asymetric  compositions, e.g. taking into account head room and look room or the rules of thirds. We can also describe unconventional framing to create unbalanced artistic composition or to show other visual elements in the scene.


\section{Shot Descriptions}
A \emph{shot} is a sequence of frames over a continuous time period. For a cohesive and coherent narration the individual shots have to be joined in a manner so as to allow the spectator to mentally recreate the story 
\cite{Cutting2016NarrativeTA}. Transition describes the progression of a shot to the subsequent one or the starting and ending of a shot. In our model we include three of the most widely used transition techniques: \emph{cut}, \emph{dissolve} and \emph{fade}. We use the simplest form of cut transition in which the two shots are played one after another. We also use the same notation to describe other types of cuts, such as cutaways in which shot A is followed by a intermittent shot with a different composition and then returns back to shot A. Dissolves and fades  are used to describe the entry or exit of a shot in which the composition either slowly appears or disappears respectively. 

As we enter the shot the initial composition can be of two types. It can either be a 'static' composition in which the actors are not performing any action or we can transition into action. In the former case, we describe the composition in the first frame of the shot. In the latter, we describe the composition in relation to the action that is being performed (figure \ref{fig:Breathless}).  This is described under the non-terminal \emph{whileEvent} which describes the first composition of the shot \emph{while} the actors are performing an \emph{Action}.  As we transition into action, the composition can also be accompanied by a camera movement. In Fig. \ref{fig:Shining} we transition directly into an actor action with camera movement. Here the camera tracks with the actor thereby resulting in maintaining a single composition in the entire action shot with no developments.

The key feature of the language is that all shots are self contained entities. Starting from a transition to the final composition, including camera and script actions, each prose storyboard sentence is independent of the previous or the next shot. A shot description can always be written and read without requiring knowledge from the previous or next shot in a movie. 


Based on the taxonomy of shots proposed by Thomson and Bowen \cite{Thompson09a},  the prose storyboard  language allows to describe precisely three main categories of shots :

\begin{itemize}
	\item A simple shot is taken with a camera that does not move or turn. Any change in the composition is from the movements of the actors in relation to the camera.	
	\item A complex shot is taken with a camera that has movements around a fixed point such as pan, tilt and zoom. We introduce camera actions pan and zoom to describe such movements. Thus the camera can pan left and right, up and down (as in a tilt) and zoom in and out.  
	\item A developing shot is taken with a moving camera. We introduce two camera actions (dolly and crane) to describe these shots. Pan and zoom are allowed during dolly and crane movements thereby creating interesting visual effects. 
\end{itemize}

Some shots consist of a single composition from beginning to end. In many cases, however, the initial composition is followed by a number of  \emph{developments}. They are of two types, \emph{continuation} or \emph{recomposition}. A  \emph{continuation} is the PSL description of actions, either actor or camera, that do not lead to a change in composition from the previous one. It can be from an \emph{action} such as speaking or looking that is important to the screenplay to mention but that does not  cause a change in composition. These are described under the \emph{event} non-terminal. Or the \emph{continuation} can be due to a \emph{follow event} in which the subjects start  to perform an action in the \emph{cue} and the camera moves with the subjects and tracks their movements so as to maintain the composition. This camera action is categorized under the \emph{camera with} non-terminal. 

The second type of \emph{development} is the  \emph{recomposition} where there is, as the name suggests, a change in composition. This change can either be due to an action performed by the \emph{subject} or by the camera or both but these two movements are not synchronized as in \emph{camera with}. These camera actions are categorized under the \emph{camera to} non-terminal. In the case of a change in composition only because of actor movements such as in simple shots (figure \ref{fig:simple_action}), the camera "holds" to the next composition.

After the initial composition, there can be any number of developments of either type. To the best of our knowledge, the prose storyboard language is the first description framework that correctly describes developing shots of arbitrary length and complexity.

For each of the three cases we propose a simplified model of a shot which consists of a sequence of compositions, cues and screen events. Cues are actor movement which trigger camera movements. They are an important construct in classical movie-making where camera movements are frequently motivated by the story. Cues are optional, which makes it possible to also describe unmotivated camera movements preceding the action, or descriptive camera movements not related to actor movements. Screen events can be actions of the camera relative to the actors, or actions of the actors relative to the camera. Screen events come in two main categories - those which change the composition ({\em events-to-composition}) and those which maintain the composition ({\em events-with-composition}). 

In our model, we can be in only one of four different states:

\begin{enumerate}
	\item Camera does not move and composition does not change.
	\item Camera does not move and composition changes due to actor movements.
	\item Camera moves and composition does not change 
	\item Camera moves and composition changes.
\end{enumerate}

In case (1), the shot can be described with a single composition. In case (2), we introduce the special verb {\em hold}  to indicate that the camera remains static while the actors move, leading to a new composition. In case (3), we use the generic construction  {\em CameraWith} (pan with, dolly with, crane with) to indicate how the camera moves to maintain the composition. In case (4), we use the generic construction  {\em CameraTo} to indicate how the cameras moves to change the composition. 

\begin{figure} 
    \centering
    \subfigure[ECU Scissors MCU Marianne front]
    {
        \includegraphics[width=0.8\columnwidth]{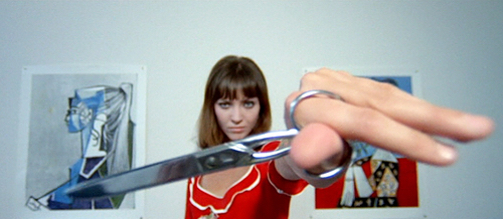}
        \label{fig:sub4}
    }
    \subfigure[CU Sanchez hands front as hands hold Bomb]
    {
        \includegraphics[width=0.8\columnwidth]{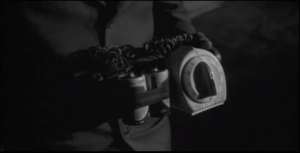}
        \label{fig:sub5}
    } 
    \caption{Compositions in ~\subref{fig:sub4} and ~\subref{fig:sub5} include inanimate objects 
    (scissors in Jean-Luc Godard's \emph{Pierrot Le Fou} and a bomb in Orson Welles's \emph{Touch Of Evil}.}

    \subfigure[MLS Father 34right screen left ELS Kane 34left screen center MS Thatcher Mother 34left screen right]
    {
        \includegraphics[width=0.8\columnwidth]{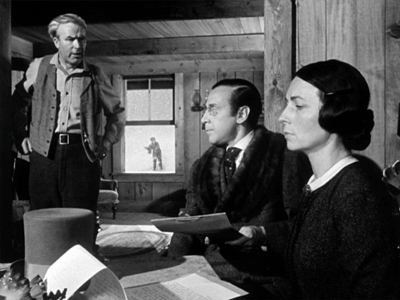}
        \label{fig:sub6}
    } 
    \subfigure[MCU Eve 34left MS Thornhill 34right Vandam 34left Leonard 34backleft]
    {
        \includegraphics[width=0.8\columnwidth]{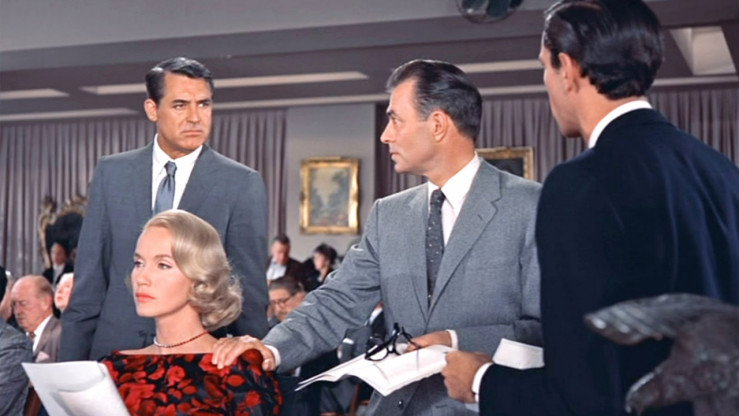}
        \label{fig:sub7}
    } 
    \caption{Complex composition  in ~\subref{fig:sub6} Orson Welles's 1941 \emph{Citizen Kane} and  ~\subref{fig:sub7} Alfred Hitchcock's 1959 \emph{North By Northwest} show  multiple actors  at different distances from the camera. They are described from left to right with their sizes indicating their depth in the composition.}
 \end{figure}

\section{Annotation results}
As a validation of the proposed language, we have manually annotated extended scenes from existing movies covering many different styles and periods.

\begin{table*}
  \caption{Annotation results: For each movie, we give the total number of annotated shots, compositions
  and developments, together with a count of the main categories of camera movement.}
  \centering
  \renewcommand{\arraystretch}{1.2}
  \begin{tabular}{|c|c|c|c|c|c|c|c|c|c|c|}
    \hline
    \textbf{Movie} &
    \textbf{Shot} &
    \textbf{Composition} &
    \multicolumn{8}{c|}{\textbf{Development}}\\
    \cline{4-11}
    &  &  & \multicolumn{3}{c|}{Continuation} & 
    \multicolumn{5}{c|}{Recomposition}\\
    \cline{4-11}
    & & & Pan & Dolly & Crane & Hold & Zoom & Pan & Dolly & Crane \\
    \hline
    Back To The Future & 41 & 69 & - & - & - & 12 & 1 & 6 & 10 & - \\
    \hline
    North By Northwest & 133 & 209 & 2 & 7 & - & 49 & 1 & 8 & 16 & -\\
    \hline
    Touch Of Evil & 1 & 40 & - & 2 & 1 & 6 & - & 1 & 18 & 13\\
    \hline
    Rope & 2 & 12 & - & 1 & - & - & - & 3 & 7 & - \\
    \hline
    \hline
    Total & 177 & 330 & 2 & 10 & 1 & 67 & 2 & 18 & 51 & 13 \\
    \hline
  \end{tabular}
  \label{tab:annotated_example}
\end{table*}

\subsection{Process of annotation}
We start the annotation process by viewing the scenes multiple times. The scene is then divided into its consisting shots which we describe in Prose Storyboard Language. Each of these sentences is matched to their corresponding keyframe in the shot via timecodes. For example, the first frame of the shot matches the initial composition in PSL. The time code for this keyframe is noted. As the shot progresses, we make a note of the subsequent compositions and their time codes. This list of PSL sentences with time codes for keyframes is written as subtitles in a word processor and saved as  SubRip(.srt) files that can be played with the annotated scene. To make the PSL sentences easier to read they are generally broken down into fragments that start with a 'from' composition and the time codes for the duration this composition lasts. Then the next PSL fragment contains the action that either the camera or actor or both perform that changes this initial composition. After this, we describe the 'to' composition that the previously mentioned action leads to. All these fragments are accompanied by their time codes. The output of this process of annotation is a time-coded PSL description of the scene in a .srt file. 

These PSL sentences are then parsed using the PSL grammar described in the AND/OR tree in Fig. \ref{fig:psl_andortree}. This parsing is done in Python using the Parsimonious toolkit. The parser is based on parsing expression grammars (PEGs) in which lexing and parsing are done at the same time. The key feature of PEGs is that it uses a prioritized choice operator "/" rather than an unordered operator "|". This means the order in which the choices are written is important (For example, in a rule 'A = a / b / c ', the parser first checks if the input matches 'a'. It only moves to the next choice if this fails).  This prioritization removes ambiguity and ensures there is only one output parse tree for a given input. A reference implementation of the PSL parser is freely available at \url{https://gitlab.inria.fr/vmurukut/psl} together with all PSL sentences mentioned in the paper.

\subsection{Annotation results}

We annotated   scenes extracted from four movies : Back To The Future by Robert Zemekis, Rope and North By Northwest by Alfred Hitchcock, and Touch Of Evil by Orson Welles. In each case, we give the original screenplay, the movie subtitled with a complete PSL description of all compositions and developments, and a storyboard with one keyframe per composition or development. Annotation results are summarized in Table \ref{tab:annotated_example} and can be found in the accompanying material available online at \url{https://team.inria.fr/anima/prose-storyboard-language/}. With 177 shots and 330 compositions, they constitute an informal validation of the expressivity and generality of the language, as well as an illustration of good practices for precisely annotating movie shots using the language.

The cafe scene in Back To The future (1985) is an interior scene with a combination of action and dialogues involving 8 main characters. Scene elements for this movie are enumerated in Fig. \ref{grammar_bttf} and appended to the PSL grammar for annotation. We are making the prose storyboard for all 41 shots in the scene available for future reference. Rope (1948), a single shot movie by Alfred Hitchcock, also shows action and dialogue between 8 characters, this time using elaborate blocking and camera movements rather than employing cuts. Scene elements for this movie are enumerated in Fig. \ref{grammar_rope}. This is challenging example for annotation, and we show examples from two extended sequences fully annotated  with PSL. Results are shown in Fig. \ref{fig:rope-sketch} and \ref{fig:rope-frames}. 

We also annotated the crop duster  scene from North By Northwest (1959) to highlight the versatility of the language in describing a complex scene in an outdoor environment involving many non human elements.  In that scene the intent of the pilot is personified in the movements of the plane. We annotated all 133 shots in this virtuoso scene with their prose storyboards, to illustrate the variety of shots used in this mostly silent scene. Scene elements for this movie are enumerated in Fig. \ref{grammar_nbnw}. Finally we annotated the long opening shot from Orson Welles's Touch Of Evil (1958) which shows a wide variety of camera movements interlaced with meticulously planned choreography for the characters resulting in a rich and dynamic visual composition. Scene elements for this movie are enumerated in Fig. \ref{grammar_touchofevil}. Despite the complexity of these scenes, they show that the prose storyboard is fairly simple to read and easy to generate.

\section{Conclusion}
We have presented a language for describing the spatial and temporal structure of movies with arbitrarily complex shots. The language can be extended in many ways, e.g. by taking into account lens choices,  depth-of-field  and lighting, and diegetic sound including speech. Future work will be devoted to the dual problems of automatically  generating movies from prose storyboards in machinima environments, and automatically describing shots in existing movies. We are also planning to extend our framework for the case of stereoscopic movies, where image composition needs to be extended to include the depth and disparity of subjects in the composition. It would also be interesting to extend the language even further for the case of panoramic video and immersive virtual reality movies. At this stage, we believe that the proposed language can be useful to extend existing approaches in intelligent  cinematography and editing towards more expressive strategies and idioms, and to bridge the gap between  real and virtual movie-making. 


\bibliographystyle{eg-alpha}
\bibliography{ProseStoryboardLite}  

\begin{thebibliography}{\uppercase{Ron21}}

\bibitem[Ron21]{Ronfard2021}
\textsc{Ronfard R.}:
\newblock {Film Directing for Computer Games and Animation}.
\newblock \emph{{Computer Graphics Forum} 40}, 2 (May 2021), 713--730.
\newblock Eurographics State of the Art Report (STAR).

\end{thebibliography}


\newcommand{\etalchar}[1]{$^{#1}$}
\begin{thebibliography}{\uppercase{CAwH{\etalchar{*}}96}}

\bibitem[ASSN03]{Akerberg2003}
\textsc{Akerberg O., Svensson H., Schulz B., Nugues P.}:
\newblock Carsim: An automatic 3d text-to-scene conversion system applied to
  road accident reports.
\newblock In \emph{Proceedings of the Tenth Conference on European Chapter of
  the Association for Computational Linguistics - Volume 2} (2003), EACL '03,
  pp.~191--194.

\bibitem[BBAG08]{BalasBAG08}
\textsc{Balas D., Brom C., Abonyi A., Gemrot J.}:
\newblock Hierarchical petri nets for story plots featuring virtual humans.
\newblock In \emph{AIIDE} (2008).

\bibitem[Bor98]{Bordwell98}
\textsc{Bordwell D.}:
\newblock \emph{On the History of Film Style}.
\newblock Harvard University Press, 1998.

\bibitem[Bra97]{Brand97}
\textsc{Brand M.}:
\newblock The "inverse hollywood problem": From video to scripts and
  storyboards via causal analysis.
\newblock In \emph{AAAI/IAAI} (1997), pp.~132--137.

\bibitem[BS16]{Braga2016}
\textsc{Braga P. H.~C., Silveira I.~F.}:
\newblock Slap: Storyboard language for animation programming.
\newblock \emph{IEEE Latin America Transactions 14}, 12 (Dec 2016), 4821--4826.

\bibitem[BvKR01]{Blackwell01}
\textsc{Blackwell L., von Konsky B., Robey M.}:
\newblock Petri net script: a visual language for describing action, behaviour
  and plot.
\newblock In \emph{Australasian conference on Computer science} (2001), ACSC
  '01.

\bibitem[CAwH{\etalchar{*}}96]{Christianson96}
\textsc{Christianson D.~B., Anderson S.~E., wei He L., Salesin D.~H., Weld
  D.~S., Cohen M.~F.}:
\newblock Declarative camera control for automatic cinematography.
\newblock In \emph{AAAI} (1996).

\bibitem[Cha12]{Chao12}
\textsc{Chao C.}:
\newblock Timing multimodal turn-taking for human-robot cooperation.
\newblock In \emph{Proceedings of the 14th ACM international conference on
  Multimodal interaction} (New York, NY, USA, 2012), ICMI '12, ACM,
  pp.~309--312.

\bibitem[CO06]{CO06}
\textsc{Christie M., Olivier P.}:
\newblock Camera control for computer graphics.
\newblock In \emph{Eurographics State of the Art Reports} (2006), Eurographics
  2006, Blackwell.

\bibitem[Cut16]{Cutting2016NarrativeTA}
\textsc{Cutting J.~E.}:
\newblock Narrative theory and the dynamics of popular movies.
\newblock In \emph{Psychonomic Bulletin \& Review} (2016).

\bibitem[DMR05]{DMR05}
\textsc{Dony R., Mateer J., Robinson J.}:
\newblock Techniques for automated reverse storyboarding.
\newblock \emph{IEE Journal of Vision, Image and Signal Processing 152}, 4
  (2005), 425--436.

\bibitem[FF06]{Friedman06}
\textsc{Friedman D., Feldman Y.~A.}:
\newblock Automated cinematic reasoning about camera behavior.
\newblock \emph{Expert Syst. Appl. 30}, 4 (May 2006), 694--704.

\bibitem[GCLR15]{Galvane2015}
\textsc{Galvane Q., Christie M., Lino C., Ronfard R.}:
\newblock {Camera-on-rails: Automated Computation of Constrained Camera Paths}.
\newblock In \emph{{ACM SIGGRAPH Conference on Motion in Games}} (Paris,
  France, Nov. 2015), {ACM}, pp.~151--157.

\bibitem[GCR{\etalchar{*}}13]{Galvane2013}
\textsc{Galvane Q., Christie M., Ronfard R., Lim C.-K., Cani M.-P.}:
\newblock {Steering Behaviors for Autonomous Cameras}.
\newblock In \emph{{MIG 2013 - ACM SIGGRAPH conference on Motion in Games}}
  (Dublin, Ireland, Nov. 2013), MIG '13 Proceedings of Motion on Games, {ACM},
  pp.~93--102.

\bibitem[GCSS06]{GoldmanCSS06}
\textsc{Goldman D.~B., Curless B., Salesin D., Seitz S.~M.}:
\newblock Schematic storyboarding for video visualization and editing.
\newblock \emph{ACM Trans. Graph. 25}, 3 (2006), 862--871.

\bibitem[GLC{\etalchar{*}}18]{Galvane2018}
\textsc{Galvane Q., Lino C., Christie M., Fleureau J., Servant F., Tariolle
  F.-l., Guillotel P.}:
\newblock Directing cinematographic drones.
\newblock \emph{ACM Trans. Graph. 37}, 3 (July 2018), 34:1--34:18.

\bibitem[GR13]{GR13}
\textsc{Gandhi V., Ronfard R.}:
\newblock Detecting and naming actors in movies using generative appearance
  models.
\newblock In \emph{CVPR} (2013).

\bibitem[GR15]{GR15}
\textsc{Gandhi V., Ronfard R.}:
\newblock {A Computational Framework for Vertical Video Editing}.
\newblock In \emph{{4th Workshop on Intelligent Camera Control, Cinematography
  and Editing}} (Zurich, Switzerland, May 2015), {Eurographics}, {Eurographics
  Association}, pp.~31--37.

\bibitem[GRCS14]{Galvane2014}
\textsc{Galvane Q., Ronfard R., Christie M., Szilas N.}:
\newblock {Narrative-Driven Camera Control for Cinematic Replay of Computer
  Games}.
\newblock In \emph{{MIG'14 - 7th International Conference on Motion in Games }}
  (Los Angeles, United States, Nov. 2014), {ACM}, pp.~109--117.

\bibitem[GRG14]{GR14}
\textsc{Gandhi V., Ronfard R., Gleicher M.}:
\newblock {Multi-Clip Video Editing from a Single Viewpoint}.
\newblock In \emph{{CVMP 2014 - European Conference on Visual Media
  Production}} (London, United Kingdom, Nov. 2014), {ACM}, p.~Article No. 9.

\bibitem[HCS96]{He96}
\textsc{He L.-w., Cohen M.~F., Salesin D.~H.}:
\newblock The virtual cinematographer: a paradigm for automatic real-time
  camera control and directing.
\newblock In \emph{Proceedings of the 23rd annual conference on Computer
  graphics and interactive techniques} (New York, NY, USA, 1996), SIGGRAPH '96,
  ACM, pp.~217--224.

\bibitem[JY05]{Jhala05}
\textsc{Jhala A., Young R.~M.}:
\newblock A discourse planning approach to cinematic camera control for
  narratives in virtual environments.
\newblock In \emph{AAAI} (2005).

\bibitem[JY06]{Jhala06}
\textsc{Jhala A., Young R.~M.}:
\newblock Representational requirements for a plan based approach to automated
  camera control.
\newblock In \emph{AIIDE'06} (2006), pp.~36--41.

\bibitem[LCL18]{Louarn18}
\textsc{Louarn A., Christie M., Lamarche F.}:
\newblock Automated staging for virtual cinematography.
\newblock In \emph{Proceedings of the 11th Annual International Conference on
  Motion, Interaction, and Games, {MIG} 2018, Limassol, Cyprus, November 08-10,
  2018} (2018), pp.~4:1--4:10.

\bibitem[LDTA17]{Leake2017}
\textsc{Leake M., Davis A., Truong A., Agrawala M.}:
\newblock Computational video editing for dialogue-driven scenes.
\newblock \emph{ACM Trans. Graph. 36}, 4 (July 2017), 130:1--130:14.

\bibitem[LG90]{Little1990SynchronizationAS}
\textsc{Little T. D.~C., Ghafoor A.}:
\newblock Synchronization and storage models for multimedia objects.
\newblock \emph{IEEE Journal on Selected Areas in Communications 8} (1990),
  413--427.

\bibitem[MJSB11]{MarkowitzKSB11}
\textsc{Markowitz D., Jr. J. T.~K., Shoulson A., Badler N.~I.}:
\newblock Intelligent camera control using behavior trees.
\newblock In \emph{MIG} (2011), pp.~156--167.

\bibitem[MRR98]{Magalhaes98}
\textsc{Magalhaes L.~P., Raposo A.~B., Ricarte I.~L.}:
\newblock Animation modeling with petri nets.
\newblock \emph{Computers and Graphics 22}, 6 (1998), 735 -- 743.

\bibitem[Mur86]{Murch86}
\textsc{Murch W.}:
\newblock \emph{In the blink of an eye}.
\newblock Silman-James Press, 1986.

\bibitem[MVW{\etalchar{*}}18]{Marti2018}
\textsc{Marti M., Vieli J., Wito\'{n} W., Sanghrajka R., Inversini D., Wotruba
  D., Simo I., Schriber S., Kapadia M., Gross M.}:
\newblock Cardinal: Computer assisted authoring of movie scripts.
\newblock In \emph{23rd International Conference on Intelligent User
  Interfaces} (2018), IUI '18, pp.~509--519.

\bibitem[Ond04]{Ondaatje04}
\textsc{Ondaatje M.}:
\newblock \emph{The Conversations: Walter Murch and the Art of Film Editing}.
\newblock Random House, 2004.

\bibitem[ORN09]{ONeillRN09}
\textsc{O'Neill B., Riedl M.~O., Nitsche M.}:
\newblock Towards intelligent authoring tools for machinima creation.
\newblock In \emph{CHI Extended Abstracts} (2009), pp.~4639--4644.

\bibitem[Pro08]{Proferes08}
\textsc{Proferes N.}:
\newblock \emph{Film Directing Fundamentals - See your film before shooting
  it}.
\newblock Focal Press, 2008.

\bibitem[RKV{\etalchar{*}}09]{RijsselbergenKVMW09}
\textsc{Rijsselbergen D.~V., Keer B. V.~D., Verwaest M., Mannens E., de~Walle
  R.~V.}:
\newblock Movie script markup language.
\newblock In \emph{ACM Symposium on Document Engineering} (2009), pp.~161--170.

\bibitem[Sal06]{Salt06}
\textsc{Salt B.}:
\newblock \emph{Moving Into Pictures.}
\newblock Starword, 2006.

\bibitem[Sal09]{Salt09}
\textsc{Salt B.}:
\newblock \emph{Film Style and Technology: History and Analysis (3 ed.)}.
\newblock Starword, 2009.

\bibitem[SMAY03]{Shen03}
\textsc{Shen J., Miyazaki S., Aoki T., Yasuda H.}:
\newblock Intelligent digital filmmaker dmp.
\newblock In \emph{ICCIMA} (2003).

\bibitem[SY06]{Seversky2006}
\textsc{Seversky L.~M., Yin L.}:
\newblock Real-time automatic 3d scene generation from natural language voice
  and text descriptions.
\newblock In \emph{Proceedings of the 14th ACM International Conference on
  Multimedia} (2006), MM '06, pp.~61--64.

\bibitem[TB09a]{Thompson09b}
\textsc{Thompson R., Bowen C.}:
\newblock \emph{{Grammar of the Edit}}.
\newblock Focal Press, 2009.

\bibitem[TB09b]{Thompson09a}
\textsc{Thompson R., Bowen C.}:
\newblock \emph{{Grammar of the Shot}}.
\newblock Focal Press, 2009.

\bibitem[VHL{\etalchar{*}}13]{Vogt2013}
\textsc{Vogt J., Haesen M., Luyten K., Coninx K., Meier A.}:
\newblock Timisto: A technique to extract usage sequences from storyboards.
\newblock In \emph{Proceedings of the 5th ACM SIGCHI Symposium on Engineering
  Interactive Computing Systems} (2013), EICS '13, pp.~113--118.

\bibitem[WPRC18]{Wu2018}
\textsc{Wu H.-Y., Pal\`{u} F., Ranon R., Christie M.}:
\newblock Thinking like a director: Film editing patterns for virtual
  cinematographic storytelling.
\newblock \emph{ACM Trans. Multimedia Comput. Commun. Appl. 14}, 4 (Oct. 2018),
  81:1--81:22.

\bibitem[Yan13]{Yannopoulos2013}
\textsc{Yannopoulos A.}:
\newblock Director{N}otation: Artistic and technological system for
  professional film directing.
\newblock \emph{J. Comput. Cult. Herit. 6}, 1 (Apr. 2013), 2:1--2:34.

\bibitem[YB08]{Ye2008}
\textsc{Ye P., Baldwin T.}:
\newblock Towards automatic animated storyboarding.
\newblock In \emph{Proceedings of the 23rd National Conference on Artificial
  Intelligence - Volume 1} (2008), AAAI'08, pp.~578--583.

\end{thebibliography}

\clearpage
\appendix


\begin{figure*}
\centering
\includegraphics[width=0.49\linewidth]{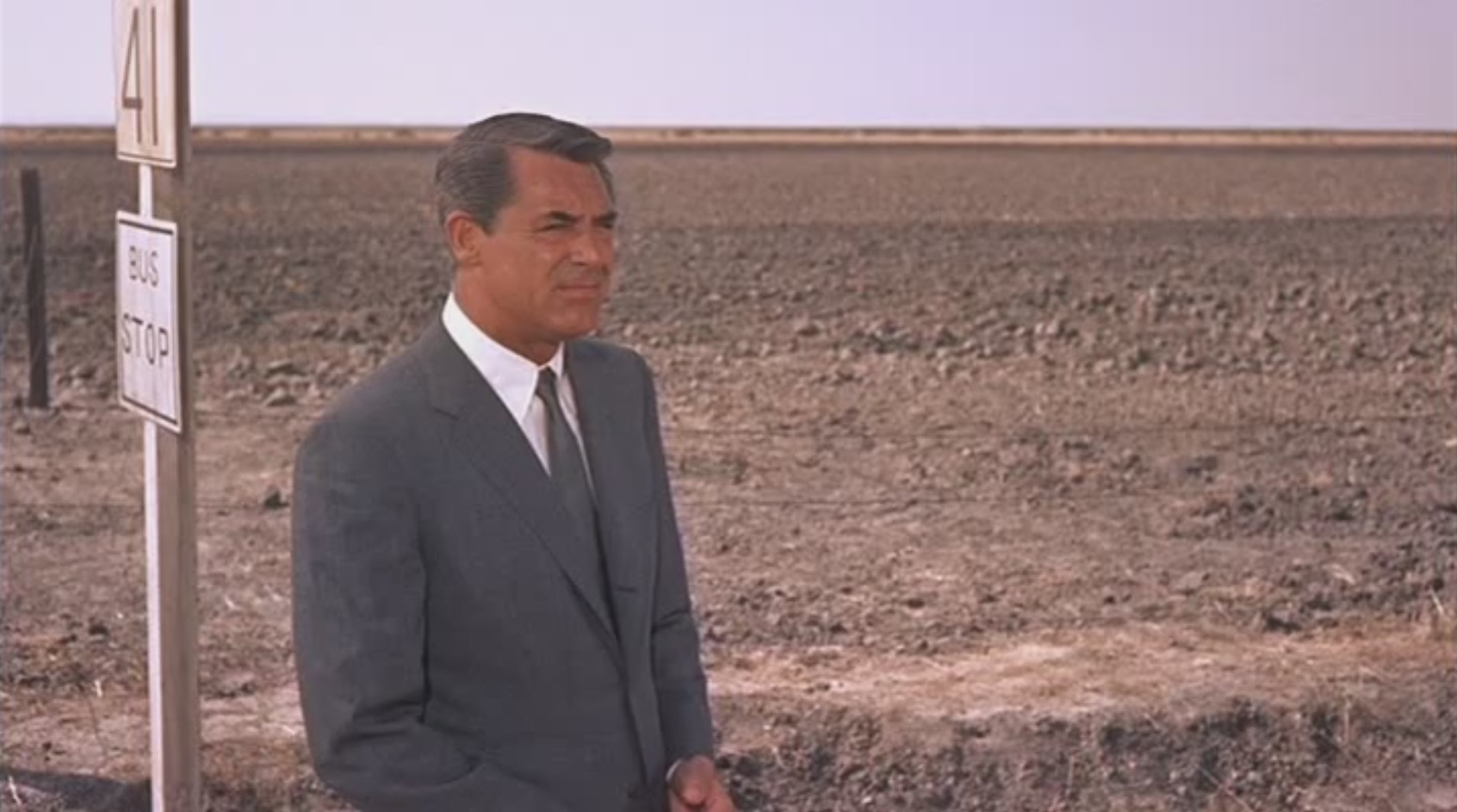}
\includegraphics[width=0.49\linewidth]{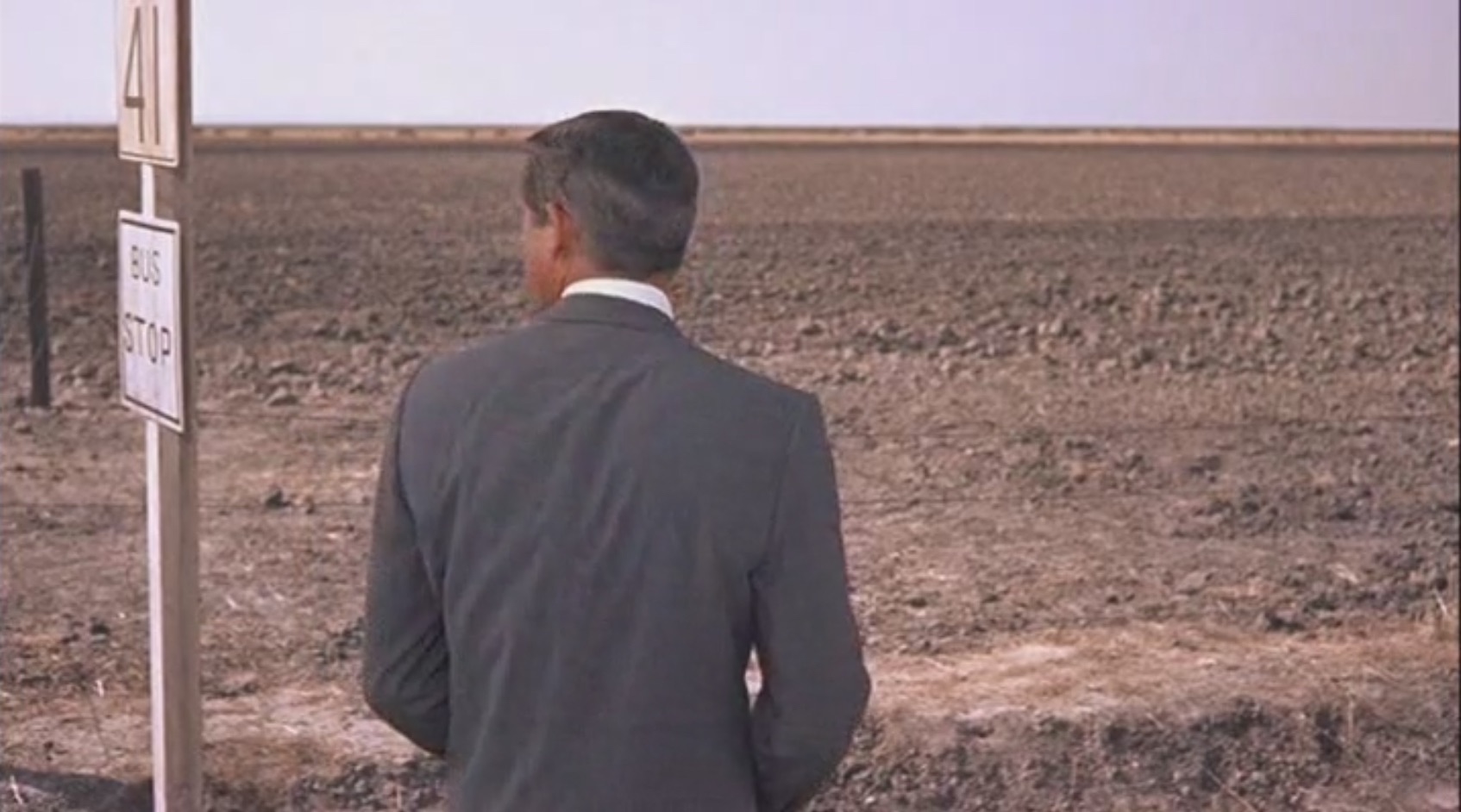}
\begin{tikzpicture}[grow'=up,scale=0.95]
\tikzset{every tree node/.style={align=center,anchor=base}}
\tikzset{level 1+/.style={level distance=2\baselineskip}}
 \tikzset{frontier/.style={distance from root=8.5\baselineskip}}
\Tree [.Shot [\qroof{MS Th 34 right}.Composition ] [. Recomposition [\qroof{then as Th turns away from camera}.Cue ]  [\qroof{hold to}.CameraTo ] [\qroof{MS Th 34backright}.Composition ]]]
\end{tikzpicture}
    
    \caption{Simple shot: with actor movement in \emph{North By Northwest}}
    \label{fig:simple_action}
\end{figure*}

\begin{figure*}
\centering
\includegraphics[width=0.49\linewidth]{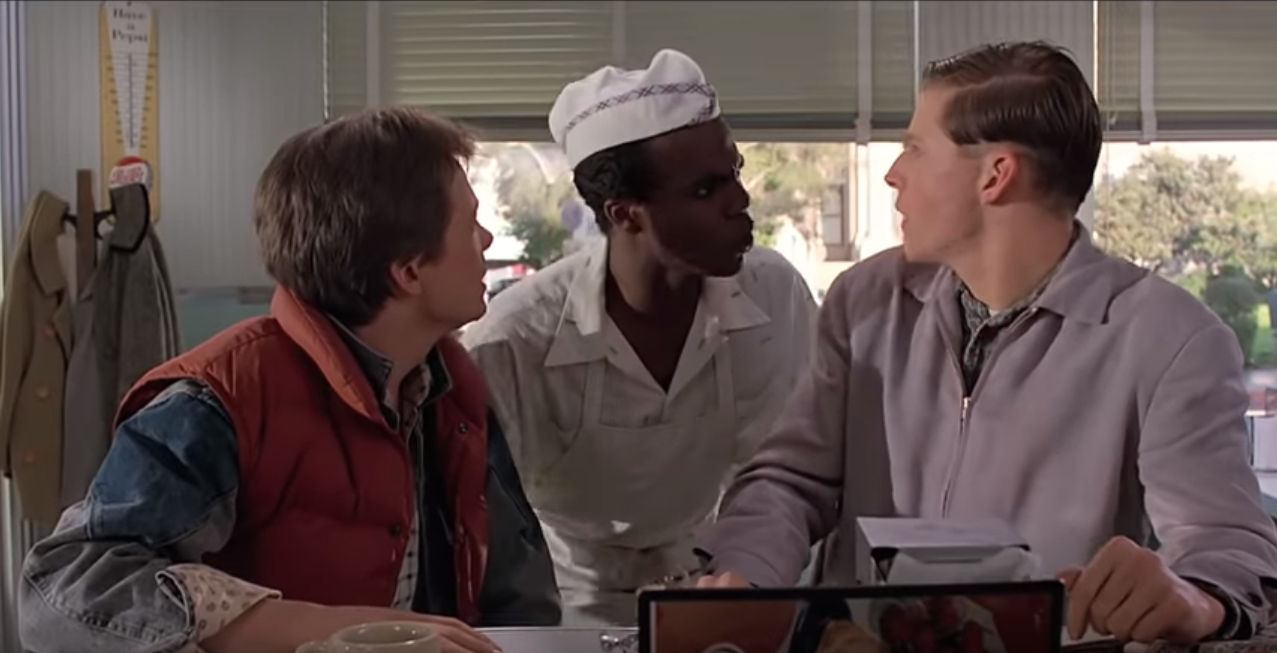}
\includegraphics[width=0.49\linewidth]{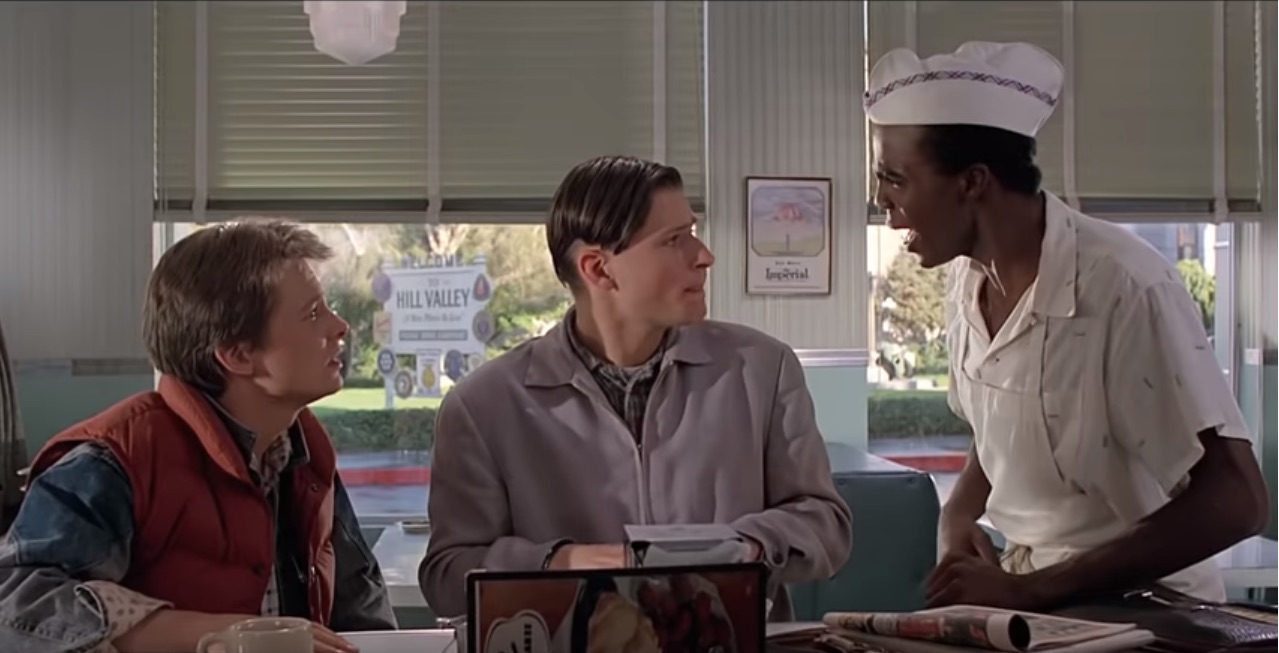}
\begin{tikzpicture}[grow'=up,scale=0.95]
\tikzset{every tree node/.style={align=center,anchor=base}}
\tikzset{level 1+/.style={level distance=2\baselineskip}}
 \tikzset{frontier/.style={distance from root=8.5\baselineskip}}
\Tree [.Shot [\qroof{MS Mar Go 34right Ge 34left }.Composition ] [.Recomposition [\qroof{then as Go crosses under Ge}.Cue ] [\qroof{dolly right to}.CameraTo ] [\qroof{MS Mar 34right Ge front Go 34left}.Composition ]]]
\end{tikzpicture}
\caption{Developing shot in \emph{Back To The Future}}
\label{fig:BTTF_2}
\end{figure*}

\begin{figure*}
    \centering
        \includegraphics[width =0.49 \linewidth]{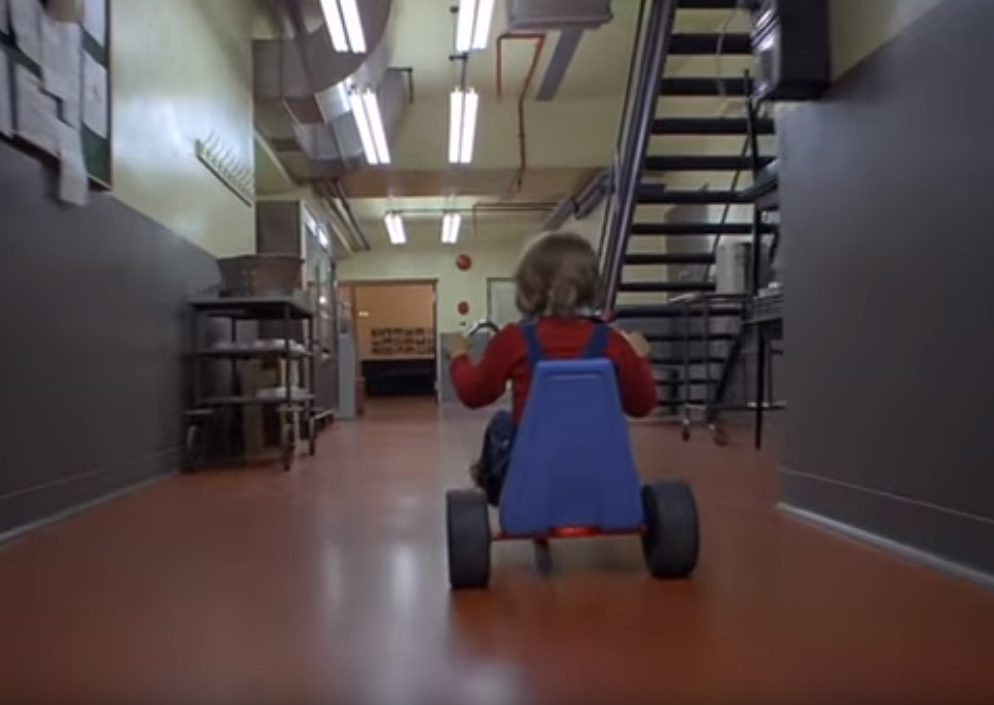}
        \includegraphics[width =0.49 \linewidth]{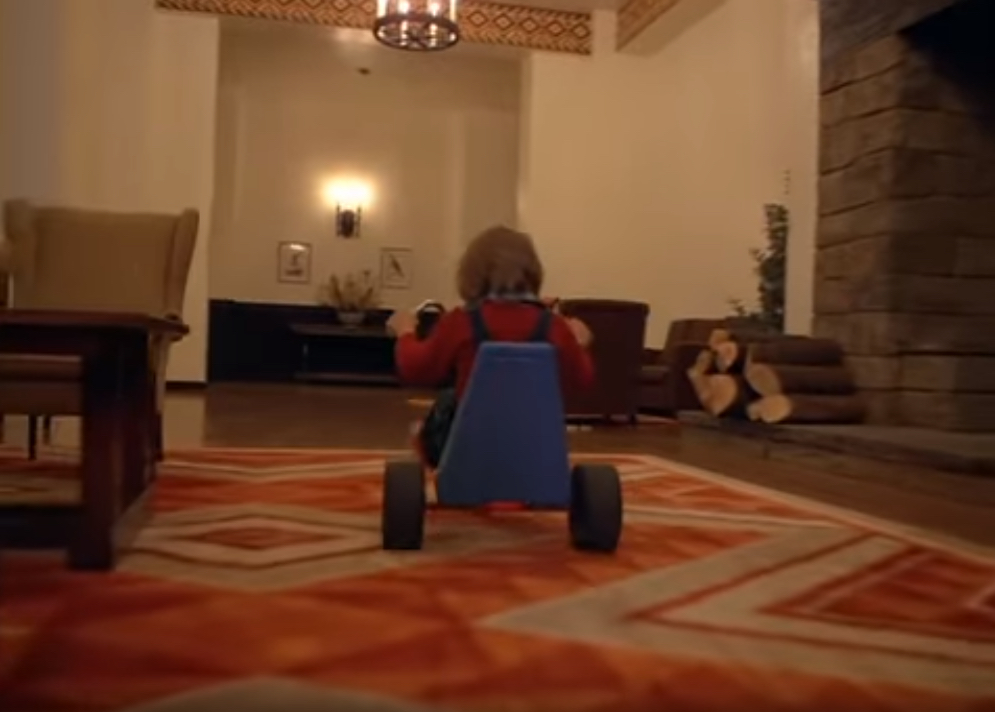}
\begin{tikzpicture}
[grow'=up,scale=0.95]
\tikzset{every tree node/.style={align=center,anchor=base}}
\tikzset{level 1+/.style={level distance=2\baselineskip}}
 \tikzset{frontier/.style={distance from root=7\baselineskip}}
\Tree [.Shot [\qroof{cut to}.Transition ] [\qroof{dolly with}.CameraWith ] [\qroof{FS Dan back}.Composition ] [\qroof{while Dan moves to screen center}.Event ]]
\end{tikzpicture}
    \caption{Developing shot: Dolly with actor in \emph{The Shining}}
    \label{fig:Shining}
\end{figure*}

\begin{figure*}
    \centering
\includegraphics[width =0.49 \linewidth]{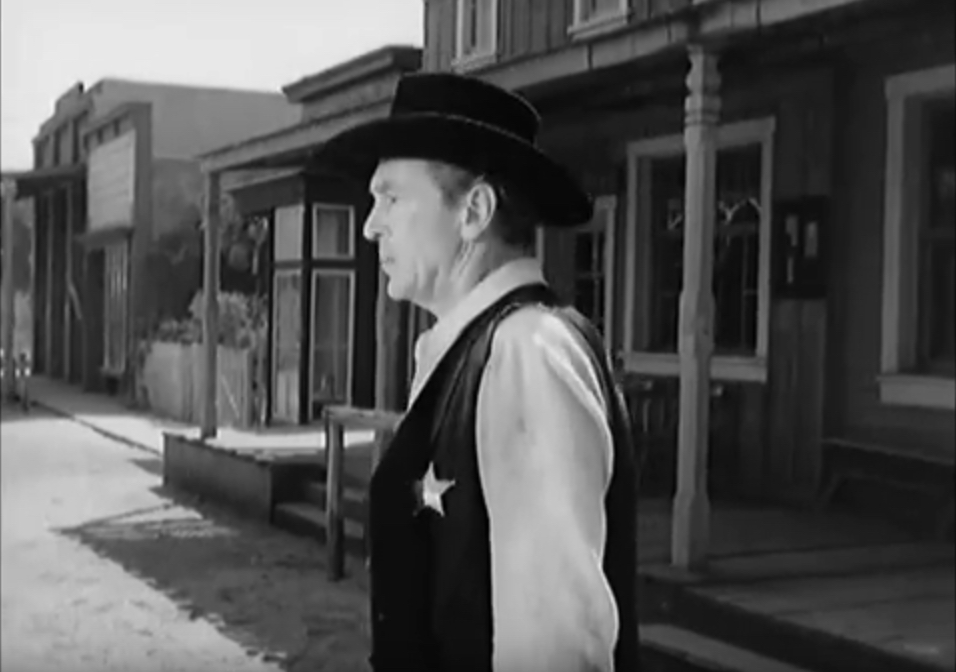}
\includegraphics[width =0.49 \linewidth]{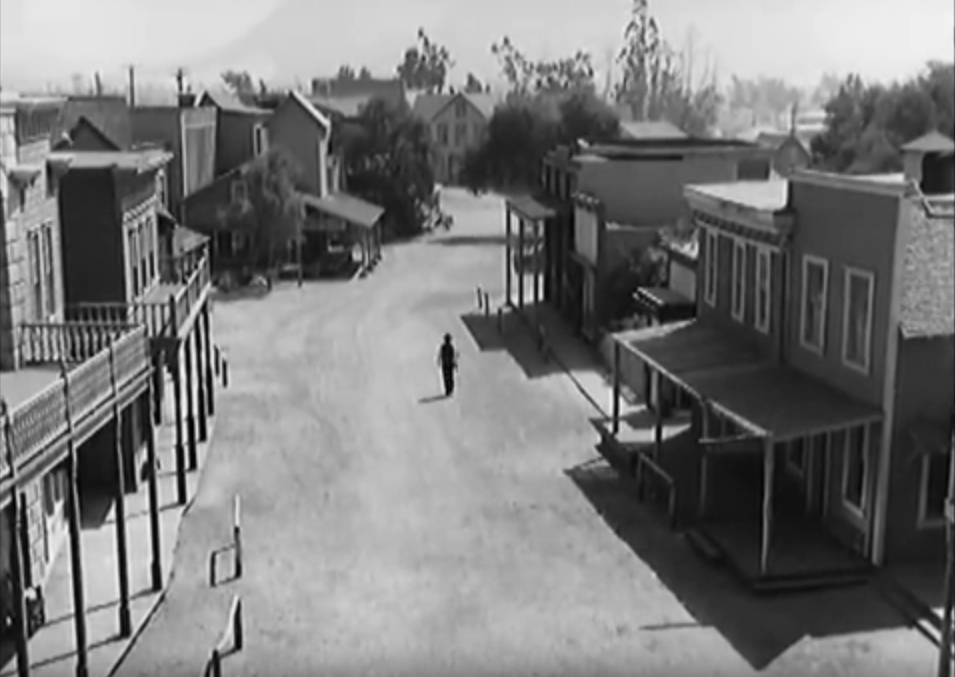}
\begin{tikzpicture}[grow'=up,scale=0.95]
\tikzset{every tree node/.style={align=center,anchor=base}}
\tikzset{level 1+/.style={level distance=2\baselineskip}}
 \tikzset{frontier/.style={distance from root=8.5\baselineskip}}
\Tree [.Shot [\qroof{MS Sh left}.Composition ] [.Recomposition [\qroof{then as Sh moves to screen top}.Cue ] [\qroof{crane up to}.CameraTo ] [\qroof{high angle ELS Sh back}.Composition ]]]
\end{tikzpicture}

    \caption{developing shot: Crane up in \emph{High Noon}}
    \label{fig:Highnoon}
\end{figure*}

\begin{figure*}
\centering
\includegraphics[width = 0.3\linewidth]{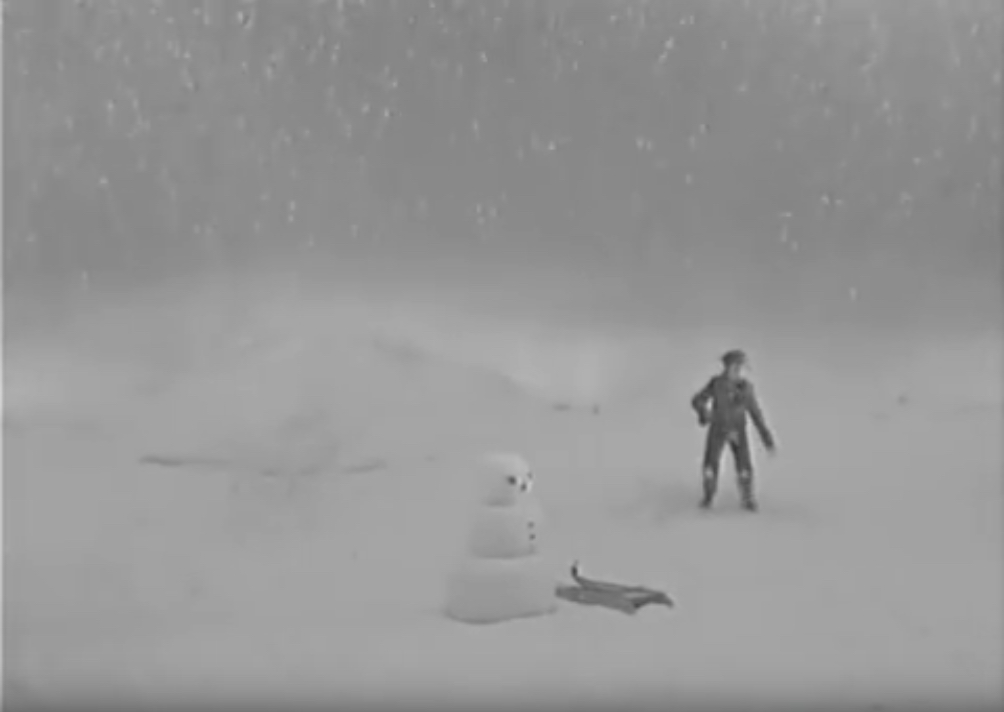}
\includegraphics[width = 0.3\linewidth]{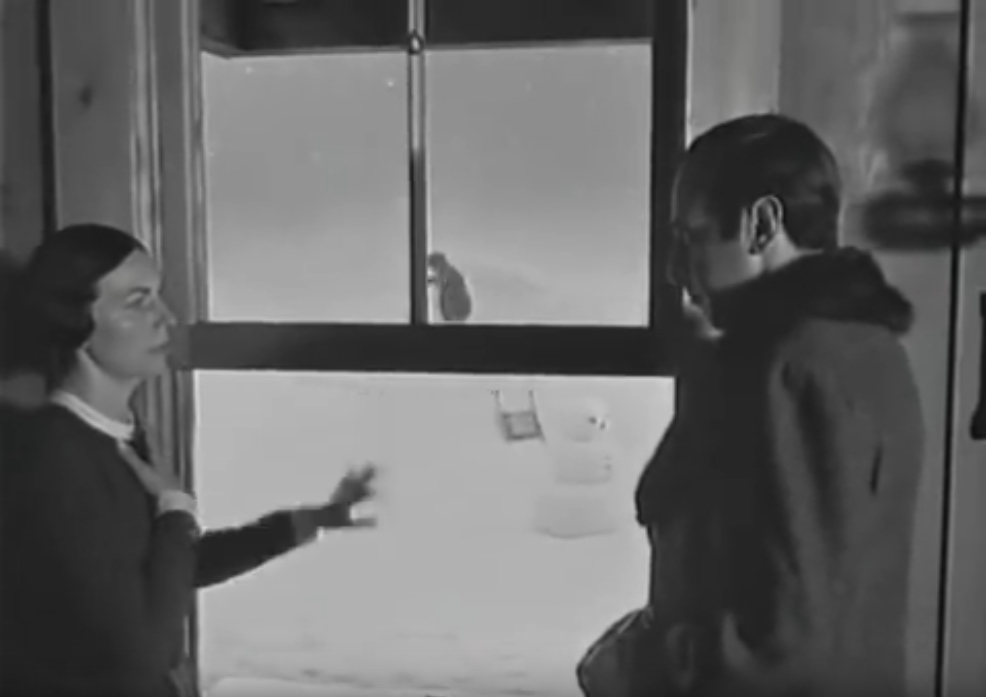}
\begin{tikzpicture}[grow'=up,scale=0.9]
\tikzset{every tree node/.style={align=center,anchor=base}}
\tikzset{level 1+/.style={level distance=2\baselineskip}}
 \tikzset{frontier/.style={distance from root=7\baselineskip}}
\Tree [.Shot... [\qroof{LS Kane}.Composition ] [.Recomposition [\qroof{then as M appears screen left}.Cue ] [\qroof{dolly out to}.CameraTo ] [\qroof{MS M right LS Kane 34backleft MS Th left}.Composition ]]]
\end{tikzpicture}

\includegraphics[width = 0.3\linewidth]{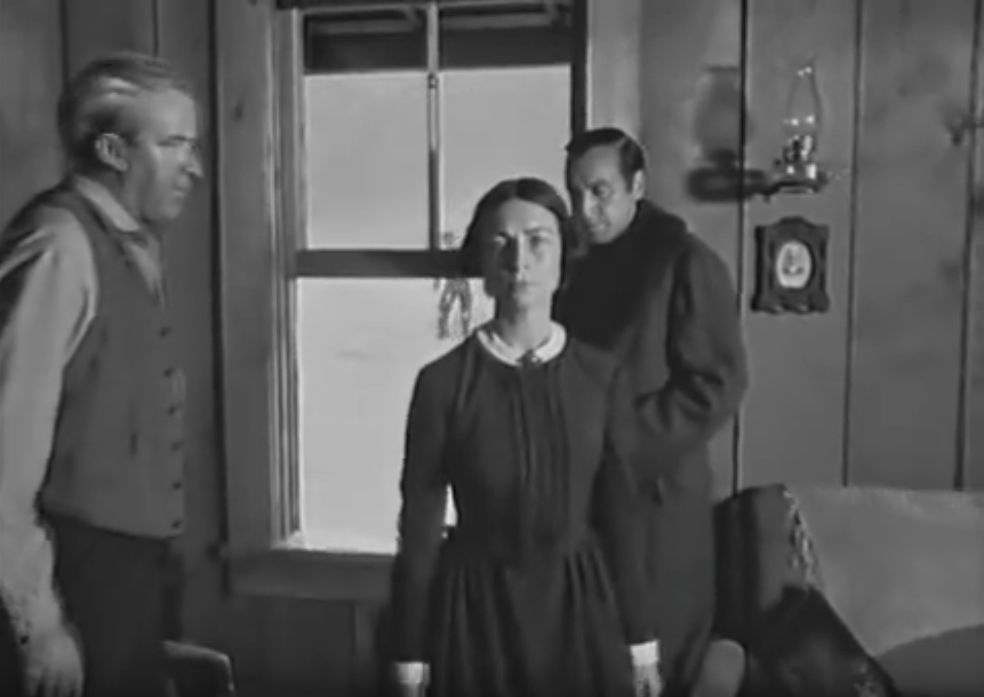}
\includegraphics[width = 0.3\linewidth]{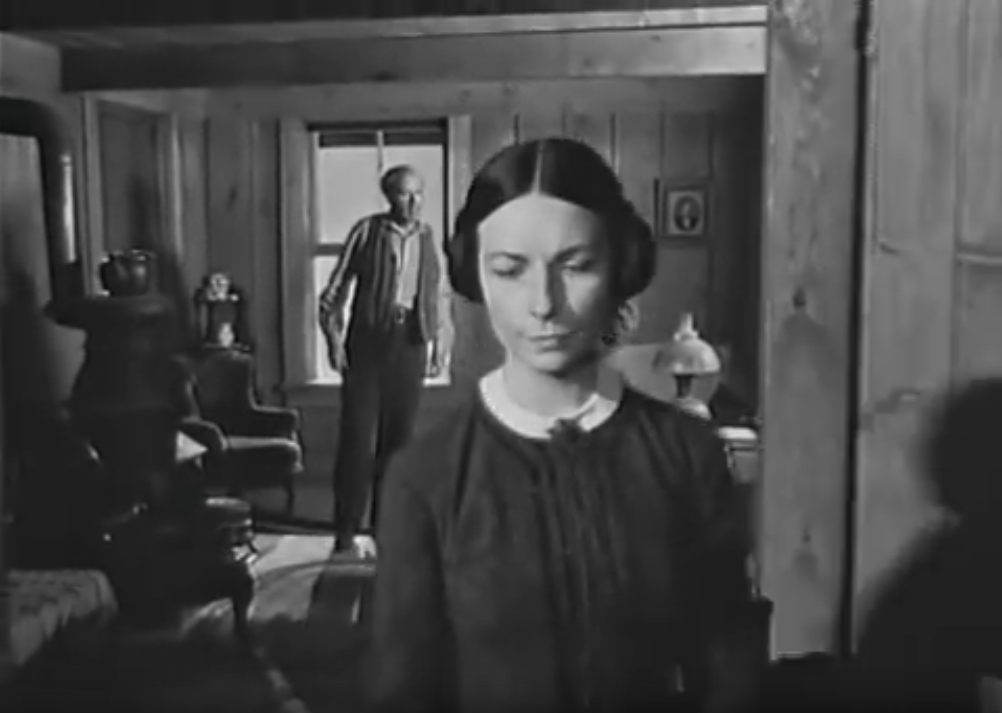}
\begin{tikzpicture}[grow'=up,scale=0.9]
\tikzset{every tree node/.style={align=center,anchor=base}}
\tikzset{level 1+/.style={level distance=2\baselineskip}}
 \tikzset{frontier/.style={distance from root=6.5\baselineskip}}
\Tree [. (continued) [.Recomposition [\qroof{then as F appears screen left and M crosses over Th}.Cue ] [\qroof{dolly out to}.CameraTo ] [\qroof{LS F MS M}.Composition ]]]
\end{tikzpicture}

\includegraphics[width = 0.3\linewidth]{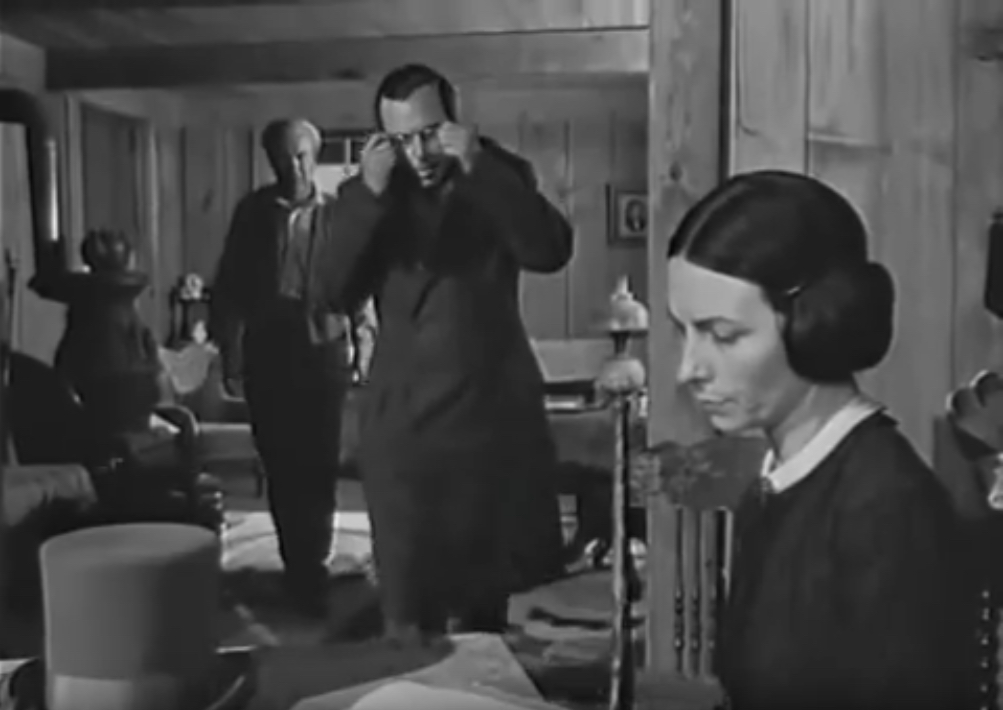}
\includegraphics[width = 0.3\linewidth]{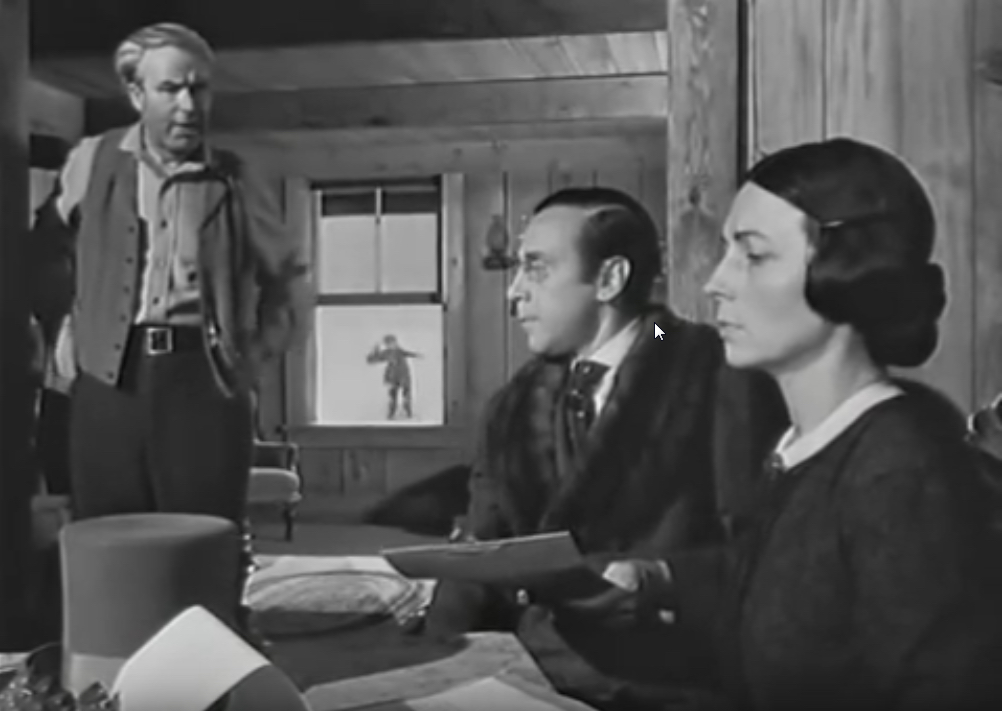}
\begin{tikzpicture}[grow'=up,scale=0.9]
\tikzset{every tree node/.style={align=center,anchor=base}}
\tikzset{level 1+/.style={level distance=2\baselineskip}}
 \tikzset{frontier/.style={distance from root=6.3\baselineskip}}
\Tree [. (continued)  [.Recomposition [\qroof{then as Th F moves to M}.Cue ] [\qroof{pan down to}.CameraTo ] [\qroof{MLS F 34right ELS Kane 34left MS Th M 34left}.Composition ]]]
\end{tikzpicture}
\caption{Developing shot with multiple actors in \emph{Citizen Kane}}
\label{fig:Kane}
\end{figure*}

\begin{figure*}
    \centering
    \includegraphics[width=\linewidth]{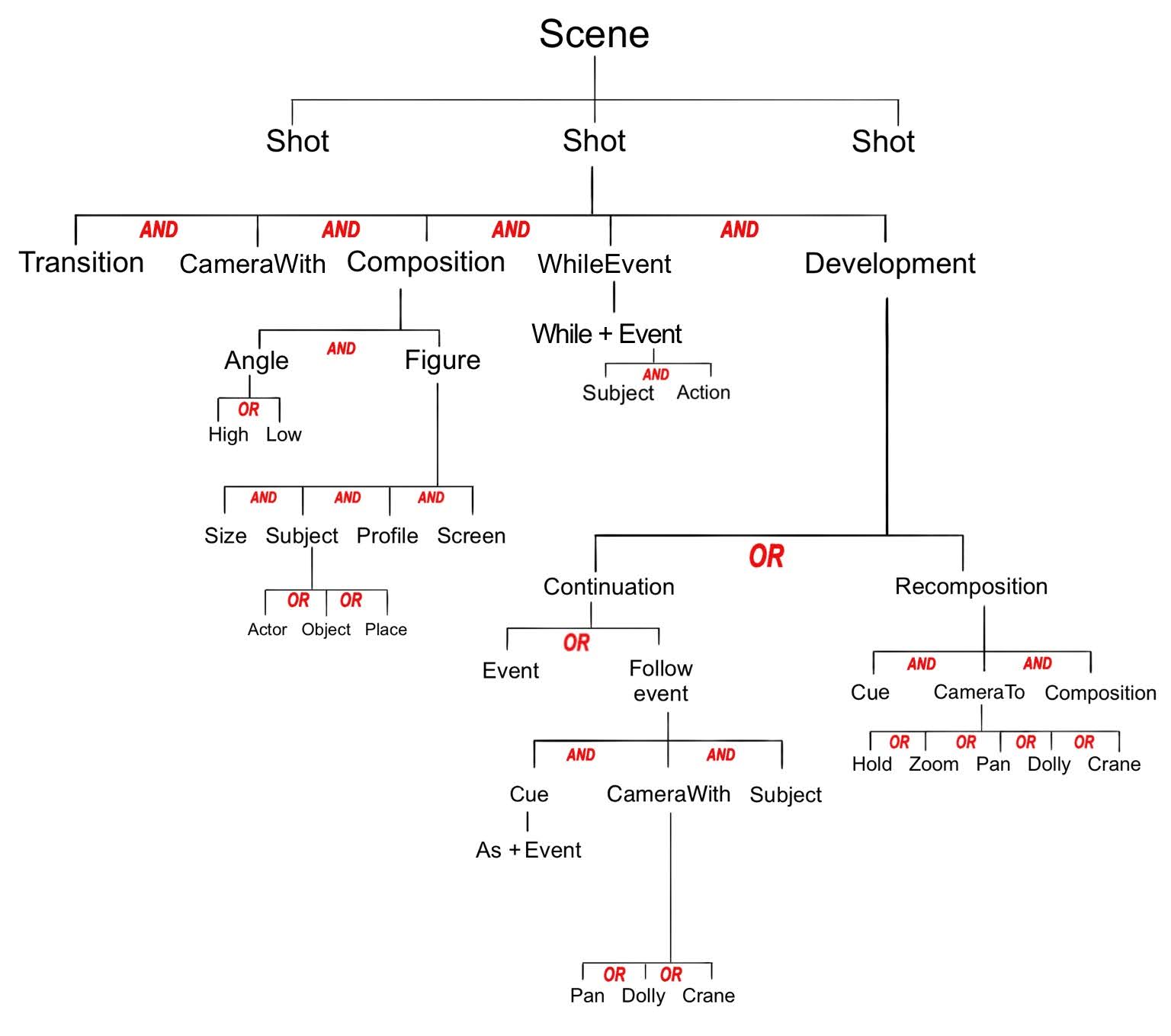}
    \caption{AND-OR tree representation of the Prose Storyboard Language grammar.}
    \label{fig:psl_andortree}
\end{figure*}

\begin{figure*} 
\centering 
\small
\begin{lstlisting}
Scene          = Shot*
Shot           = Transition? _ CameraWith? _ Composition? _ WhileEvent?
                  _ (Development?)*
Transition     = ("cut to" / "dissolve to" / "fade in to")
Development    = "then"? _ (Recomposition / Continuation)
Continuation   = Event / FollowEvent
Recomposition  = Cue? _ CameraTo _ Composition
WhileEvent     = "while"_ Event
FollowEvent    = Cue? _ CameraWith _ Agent _ ("and"? _ Agent)*
Cue            = "as" _ Event _ ("and"? _ Event)*
Composition    = (Angle? _ Figure)*
Figure         = Size? _ Subject _ Profile? _ Screen?
Agent          = (Actor / Object) _ (Actor / Object)*
Subject        = Actor / Object / Place
Angle          = "low angle" / "high angle"
Size           = "ECU" / "CU" / "MCU" / "MS" / "MLS" / "FS" / "LS" / "ELS"
Profile        = "left" / "right" / "front" / "back" / "34left" / "34right"
                    / "34backleft" / "34backright"
Screen         = "screen" _ ("top" / "bottom")? _ ("left" / "center"/ "right")?
CameraWith     = Speed? _ (Pan / Dolly / Crane)  _ "with"
CameraTo       = (Hold /(Speed? _ (Pan / Dolly / Crane / Zoom))) _ "to"
Hold           = "hold"
Pan            = "pan"   _ ("left" / "right" / "up" / "down")?
Dolly          = "dolly" _ ("left" / "right" / "in" / "out")?
Crane          = "crane"  _ ("up" / "down") _ ("left"/"right")?
Zoom           = "zoom" _ ("in" / "out")
Speed          = "slow" / "quick"
Enter          = "enters"  _ Screen? _ Place?
Exit           = "exits" _ Screen? _ Place?
Look           = "looks" _ "at"? _ (Subject / Screen)
Move           = "moves" _ "to" _ (Subject / Screen )
Speak          = ("speaks" / ("says" _ String)) _ ("to" _ Subject)?
Use            = "uses" _ Object
Cross          = "crosses" _ ("over" / "under") _ Subject
Touch          = "touches" _ Subject
React          = "reacts to" _ Subject
Turn           = "turns" _ ("left" / "right"
                    / "towards camera" / "away from camera")
Stop           = "stops" _ ("at" / "near") _ Subject
Appear         = "appears" _ (Screen / ("from behind" _ Subject))
Disappear      = "disappears" _ (Screen/ ("behind" _ Subject))
Action         = Enter / Exit / Look / Move / Speak / Use / Cross / Touch
                   / React / Turn / Stop / Appear / Disappear
Event          = Agent _ Action
\end{lstlisting}
\caption{Grammar of the prose storyboard language in the Parsing Expression Grammar (PEG) format.}
\label{grammar}
\end{figure*}

\begin{figure*} 
\centering 
\begin{lstlisting}
Actor            = "Marty" / "George" / "Biff" / "Lou" / "Goldie" 
                   / "Match" / "Skinhead" / "hands" / "3D"
Object           = "Coffee"/ "Bar" / "Car"
Place            = "Cafe"


\end{lstlisting}
\caption{Script elements for Back To The Future.}
\label{grammar_bttf}
\end{figure*}

\begin{figure*} 
\centering 
\begin{lstlisting}
Actor            = "Brandon" / "Philip" / "Atwater" / "Janet" / 
                    "Kentley" / "Kenneth" / "Rupert"  / "Wilson" 
Object           = "Glass" 
Place            = "Salon" / "Dining room" / "Kitchen"


\end{lstlisting}
\caption{Script elements for Rope.}
\label{grammar_rope}
\end{figure*}

\begin{figure*} 
\centering 
\begin{lstlisting}
Actor            = "Thornhill" / "MBS" / "Plane" / "TD1" / "TD2" / "Farmer" /
                   "BC Driver" / "BC Woman" / "BC Man"
Object           = "Bus" / "White car" / "Limo" / "Truck" / "Blue car" /
                   "Green bus" /"Bluewhite car" / "Oil truck" / "Pickup"
                    / "Brown car" 
Place            = "Arid plot 1" / "Arid plot 2" / "Arid plot 3" /
                   "Corn field" / "Highway" / "Dirt road"

\end{lstlisting}
\caption{Script elements for North By Northwest.}
\label{grammar_nbnw}
\end{figure*}


\begin{figure*} 
\centering 
\begin{lstlisting}
Actor            = "Mike" / "Susan" / "Linnekar" / "Blonde" / "Sanchez" / 
                    "Immigration official" / "Customs official" 
Object           = "Car" / "Bomb" / "Building" /"Checkpost" 
Place            =  "Border control" / "Main street" / "Parking lot" 
                    / "Left side street" 
 
\end{lstlisting}
\caption{Script elements for Touch Of Evil.}
\label{grammar_touchofevil}
\end{figure*}

\begin{figure*}
\centering
\includegraphics[height = 0.30\textheight]{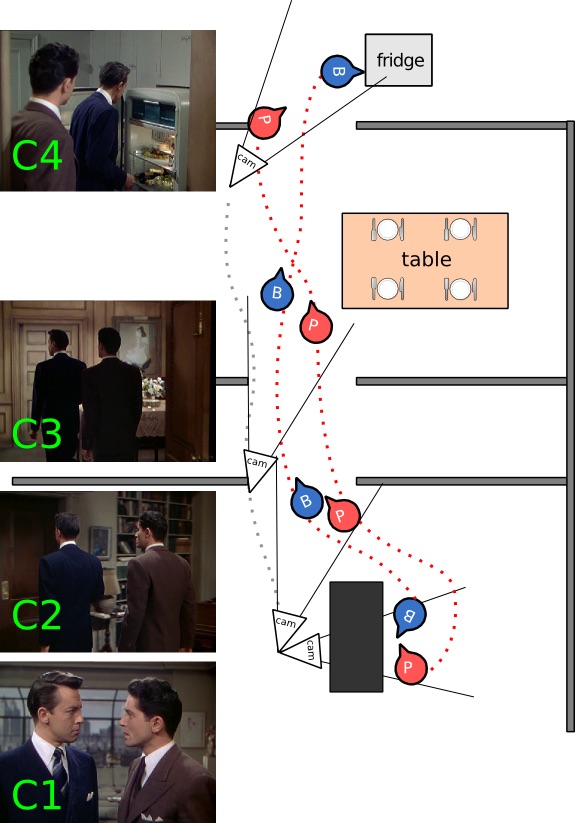}
\includegraphics[height = 0.30\textheight]{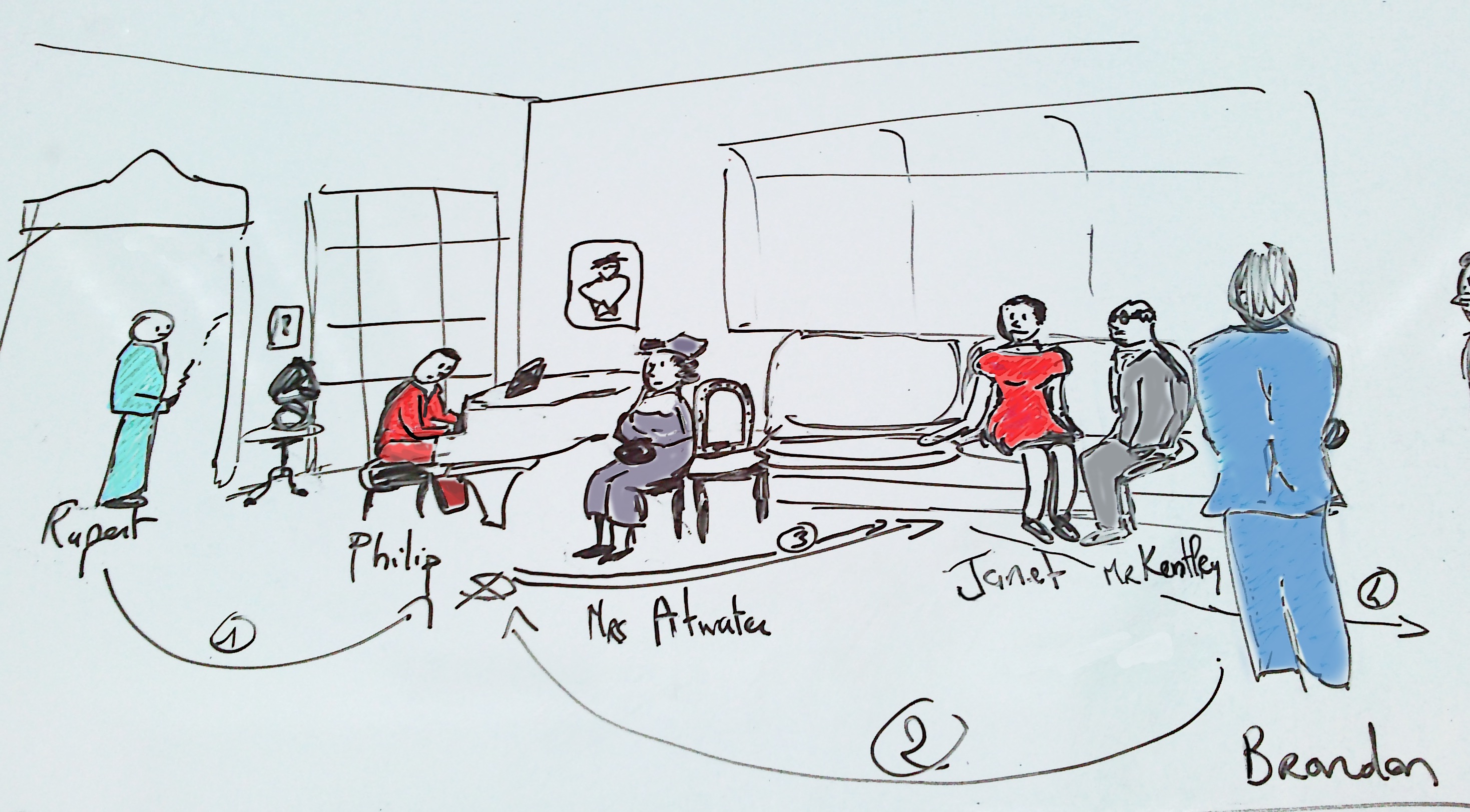}
\caption{Sketch storyboards for two sequences in Alfred Hitchcock's Rope, see Fig.\ref{fig:rope-frames} below.}
\label{fig:rope-sketch}
\end{figure*}

\begin{figure*}
    \centering
    \includegraphics[width =\linewidth]{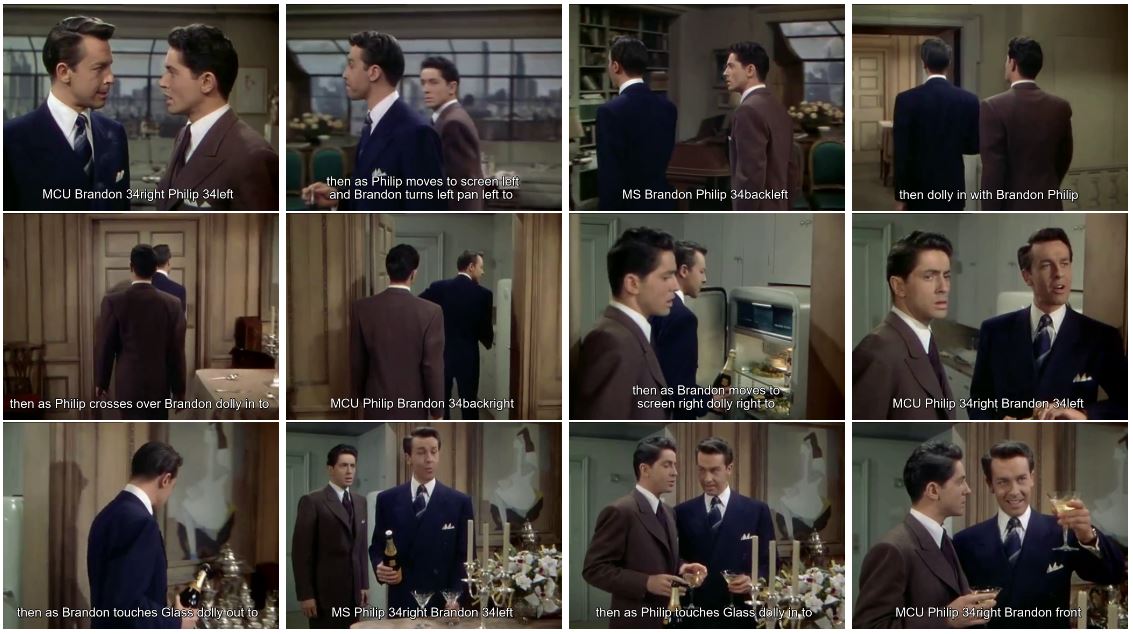}
    \includegraphics[width =\linewidth]{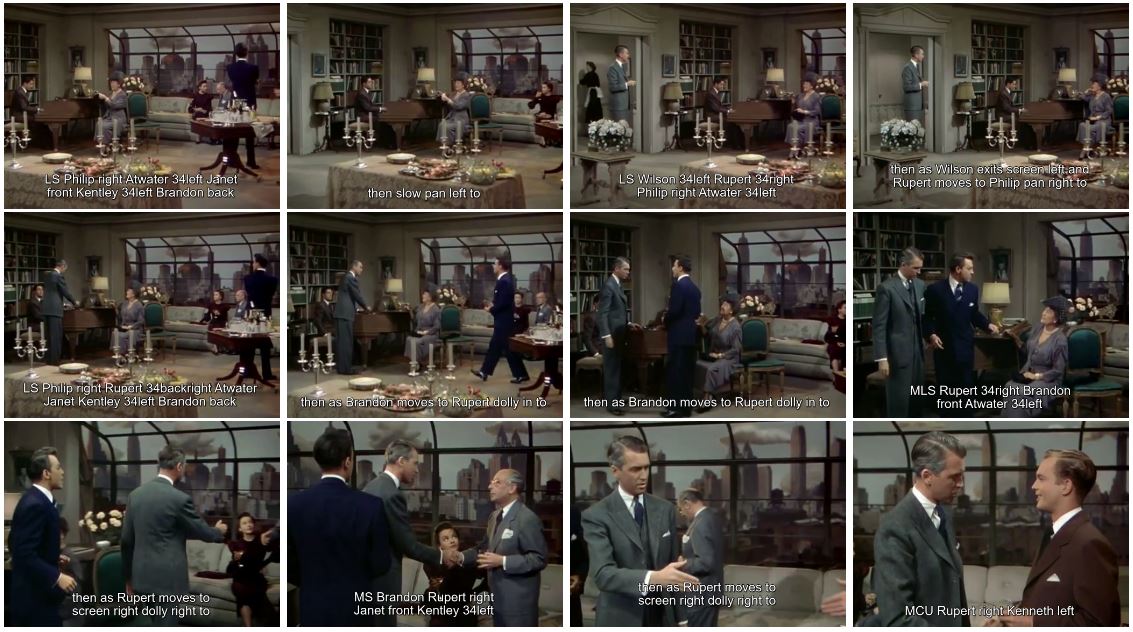}
    \caption{Prose storyboard language annotations of two extended sequences from the movie \emph{Rope}. Top three rows: First sequence from 06:55 to 08:22. Bottom three rows: Second sequence from 10:00 to 12:00.}
        \label{fig:rope-frames}
\end{figure*}

\end{document}